\documentclass[12pt]{iopart}

\usepackage{iopams}  

\newcommand{\STEREO}{\textsc{Stereo}}

\newcommand{\U}{$^{235}$U}

\usepackage[T1]{fontenc}
\usepackage{graphicx}
\usepackage{xcolor}
\usepackage{verbatim}
\usepackage{hyperref}
\usepackage{multirow}
\usepackage{lineno}
\modulolinenumbers[5]

\usepackage[normalem]{ulem}

\hypersetup{
    colorlinks=true,
    linkcolor=magenta,
    filecolor=red,      
    urlcolor=blue,
    pdftitle={\STEREO{} shape paper},
}

\begin{document}

\title[]{First antineutrino energy spectrum from \U{}~fissions with the \STEREO{} detector at ILL}

\makeatletter
\def\@fnsymbol#1{\ensuremath{\ifcase#1\or^* \or ^\dagger\or ^{\dagger\dagger}\or ^\ddagger \or ^{\ddagger\ddagger} \or ^*\or ^\mathsection\or ^\mathparagraph\or ^{**} \else\@ctrerr\fi}}
\makeatother
   

\newcommand{\Missing}[0]{\textit{\textcolor{red}{\textbf{>>???<<}}}}
\newcommand{\Replace}[2]{\textcolor{orange}{\textbf{\sout{#1}}}\textcolor{blue}{\textbf{~#2}}}
\newcommand{\Add}[1]{\textcolor{blue}{\textbf{#1}}}
\newcommand{\ValidatedReplace}[2]{#2}
\newcommand{\ValidatedAdd}[1]{#1}
\newcommand{\Comment}[1]{\textbf{\textit{\textcolor{red}{>>(#1)<<}}}}


\author{H.~Almaz\'an$^1$\footnote{Now at Donostia International Physics Center (DIPC), Paseo Manuel Lardizabal, 4, 20018 Donostia-San Sebastian, Spain}, L.~Bernard$^3$\footnote{Now at Ecole Polytechnique, CNRS/IN2P3, Laboratoire  Leprince-Ringuet, 91128 Palaiseau, France}, A.~Blanchet$^4$\footnote{Now at LPNHE, Sorbonne Universit\'e, Universit\'e de Paris, CNRS/IN2P3, 75005 Paris, France}, A.~Bonhomme$^{1,4}$, C.~Buck$^1$, P.~del~Amo~Sanchez$^2$, I.~El~Atmani$^4$\footnote{Now at Hassan II University, Faculty of Sciences, A\"in Chock, BP 5366 Maarif, Casablanca 20100, Morocco}, L.~Labit$^2$, J.~Lamblin$^3$, A.~Letourneau$^4$, D.~Lhuillier$^4$, M.~Licciardi$^3$\footnote{\href{mailto:licciardi@lpsc.in2p3.fr}{\tt licciardi@lpsc.in2p3.fr}}, M.~Lindner$^1$, T.~Materna$^4$, H.~Pessard$^2$, J.-S.~R\'eal$^3$, J.-S.~Ricol$^3$, C.~Roca$^1$, R.~Rogly$^4$, T.~Salagnac$^3$\footnote{Now at Institut de Physique Nucl\'eaire de Lyon, CNRS/IN2P3, Univ. Lyon, Universit\'e Lyon 1, 69622 Villeurbanne, France}, V.~Savu$^4$, S.~Schoppmann$^1$\footnote{Now at University of California, Department of Physics, Berkeley, CA 94720-7300, USA and Lawrence Berkeley National Laboratory, Berkeley, CA 94720-8153, USA}, V.~Sergeyeva$^2$\footnote{Now at Institut de Physique Nucl\'eaire Orsay, CNRS/IN2P3, 15 rue Georges Clemenceau, 91406 Orsay, France}, T.~Soldner$^5$, A.~Stutz$^3$, M.~Vialat$^5$}

\address{$^1$ Max-Planck-Institut f\"ur Kernphysik, Saupfercheckweg 1, 69117 Heidelberg, Germany}
\address{$^2$ Univ.~Grenoble Alpes, Universit\'e Savoie Mont Blanc, CNRS/IN2P3, LAPP, 74000 Annecy, France}
\address{$^3$ Univ.~Grenoble Alpes, CNRS, Grenoble INP, LPSC-IN2P3, 38000 Grenoble, France}
\address{$^4$ IRFU, CEA, Universit\'e Paris-Saclay, 91191 Gif-sur-Yvette, France}
\address{$^5$ Institut Laue-Langevin, CS 20156, 38042 Grenoble Cedex 9, France}

\vspace{10pt}
\begin{indented}
\item[]\STEREO{} Collaboration, \url{http://www.stereo-experiment.org}
\end{indented}

\begin{abstract}
This article reports the measurement of the \U-induced antineutrino spectrum shape by the \STEREO{} experiment. 43'000 antineutrinos have been detected at about 10 m from the highly enriched core of the ILL reactor during 118 full days equivalent at nominal power. The measured inverse beta decay spectrum is unfolded to provide a pure \U{}~spectrum in antineutrino energy. A careful study of the unfolding procedure, including a cross-validation by an independent framework, has shown that no major biases are introduced by the method. A significant local distortion is found with respect to predictions around $E_\nu \simeq 5.3$~MeV. A gaussian fit of this local excess leads to an amplitude of $A = 12.1 \pm 3.4\%$ (3.5$\sigma$).
\end{abstract}

%
%
\submitto{\jpg}

\vspace{1cm}
%
%
%


\section{\label{scn:introduction}Introduction}

The electron antineutrino spectra emitted by nuclear reactors were measured with rather high statistics in several experiments at a baseline range from $\sim 10$~m~\cite{Declais:1994su, Ko:2016owz, Ashenfelter:2018jrx} up to $\sim 1$~km~\cite{DC2020, Seo:2016uom, An:2016srz}. Anomalies with respect to the predicted antineutrino reference spectra were found, concerning the emitted absolute flux rate as well as the spectral shape. The first anomaly, known as the Reactor Antineutrino Anomaly (RAA), corresponds to a deficit of $\sim 6$\,\% on the total flux measured as compared to the predicted rate~\cite{Mention:2011rk}. The statistical significance of the RAA is reported to be up to $2.8\,\sigma$~\cite{Gariazzo:2017fdh}, depending on the choice of the prediction model~\cite{Berryman:2019hme}. The second anomaly is a shape distortion mainly expressed as a characteristic bump structure at an antineutrino energy of $E_{\nu}\!\sim\!6~\mathrm{MeV}$. Whether these differences hint towards unaccounted antineutrino physics or are introduced by the computational methods, theoretical assumptions or incomplete data inputs is still under debate. A combination of both is possible, as well. It is also not yet fully clarified if the two anomalies are connected or caused by independent effects.

The reactor antineutrino spectra can be predicted applying different methods. The first one is based on a conversion scheme using high-precision electron spectrum measurements at the BILL spectrometer at ILL~\cite{Schreckenbach:1985ep, Hahn:1989zr}. In this experiment, each of three isotopes \U, ${}^{239}$Pu and ${}^{241}$Pu were exposed to thermal neutrons to undergo fission and create $\beta$-instable fission fragments. Later, the electron spectrum of ${}^{238}$U was measured in an independent setup by fast neutron bombardment of ${}^{238}$U-foils~\cite{Haag:2013raa}. The commonly used antineutrino reference spectra from Mueller et al.~\cite{Mueller:2011nm} or Huber~\cite{Huber:2011wv} compute the antineutrino spectra of \U{}, ${}^{239}$Pu and ${}^{241}$Pu with a conversion technique starting from these measured $\beta$-spectra. In an updated conversion method calculation, forbidden decays are included via nuclear shell model calculations~\cite{Hayen:2018uyg}. This leads to an enhancement of the antineutrino flux at energies above 4~MeV going into the direction of the observed shape anomaly. At the same time, it leads to an increase of the total predicted antineutrino flux enhancing the measured absolute rate deficit. 

In another approach, the summation methods or \textit{ab initio} calculations, the antineutrino spectra are calculated using purely information contained in nuclear data bases and theoretical inputs~\cite{King:1958,Estienne:2019ujo}. Whereas hundreds of fission fragments, many of them with more than 10 different $\beta$-decay branches, contribute to the overall antineutrino spectrum, only few appear to be relevant above $\sim$ 5~MeV. Some of the predictions based on summation methods show better agreement with experimental data concerning the antineutrino rate as well as the spectral shape. However, the theoretical knowledge and experimental data on many of the involved isotopes is rather limited. In regions with dominant contributions of strongly populated fission products with sharp cutoffs in their antineutrino spectra (> 5~MeV), deviations from the smooth parametrization in the Huber or Mueller models might occur~\cite{Sonzogni:2017voo}. In general, the uncertainties on summation spectra are known to be sizeable and the shape strongly depends on the used fission yield databases. Those should be revised and improved~\cite{Hayes:2015yka}, since biases in the databases could introduce spectral structures similar to the ones observed.

Several currently running short baseline reactor antineutrino experiments, including \STEREO{}~\cite{Almazan_2020}, are testing if the overall rate deficit could be linked to neutrino flavor oscillations into a light sterile state~\cite{Ko:2016owz, Alekseev:2018efk, Ashenfelter:2018iov, Serebrov:2018vdw, Abreu:2018pxg}. Such an oscillation phenomenon could explain the missing electron antineutrino flux at the position of the detector, but not the observed shape distortion in the antineutrino spectra. Another open question to be resolved is whether the rate and shape anomalies are solely created by a single fissile isotope as \U{}~or if other isotopes contribute with similar or even larger extent. In this regard, a comparison between an experiment as \STEREO{} performed at a reactor highly enriched in \U{}~(HEU) with an experiment at a commercial power reactor using low uranium enrichment (LEU) might give further insights~\cite{ratePaper,Buck:2015clx}. So far, most experiments were operated at LEU reactors for which up to 60\,\% of the emitted antineutrino flux is created by beta decays of fission products other than \U.

The shape anomaly has been revealed by most experiments as an excess of events as compared to the predictions in the region of 5-7 MeV, after correcting the normalization by the observed rate deficit from the RAA. The local significance of this distortion is in several cases more than $3\,\sigma$. However, the event excess varies in magnitude and the peak positions are sometimes slightly shifted. Although there is widespread expectation that the spectral distortion is linked to inaccurate modeling of the antineutrino spectra, alternative explanations related to the detector performances exist. In particular, calibration data in this energy region is often limited and a common bias in the non-linearity models could, in principle, also create such distortions~\cite{Mention:2017dyq}. The good agreement between the measured spectrum in the Bugey-3 data~\cite{Declais:1994su} with the Huber and Mueller models and its discrepancy to other spectra remain a puzzle.  

Now, new experimental data from different reactor types could bring valuable insights into the nature of the reactor shape distortion. The computation of the predicted spectrum is simplified in research reactors using HEU fuel due to the strongly reduced impact of burn-up effects. A shape-only analysis of the PROSPECT data~\cite{andriamirado2020improved} finds in the region of 5-7 MeV (antineutrino energy) a behavior compatible with the high statistics Daya Bay data~\cite{An:2016srz}. This analysis disfavors the Huber's prediction and the scenario of a pure \U{} origin of the excess with more than $2\sigma$ significance. A recent analysis of the Daya Bay experiment allowed to extract antineutrino spectra individually for different fission isotopes. A slightly weaker distortion was observed for ${}^{235}$U as compared to ${}^{239}$Pu, the second dominant isotope contributing to the thermal power production of a typical LEU reactor~\cite{Adey:2019ywk}. 

In this article, the shape analysis of the \STEREO{} experiment is discussed providing a high statistics sample. It aims at providing an accurate and model-independent \U-induced IBD yield spectrum in antineutrino energy. The structure of this article is as follows. In section~\ref{scn:Experiment} we present the experimental site at ILL and the \STEREO{} detector. The detector response, its tuning in the simulation and related systematic uncertainties, are presented in section~\ref{scn:Response}. Section~\ref{scn:Data} details the event selection, background modelling and extraction of the antineutrino data spectrum. We introduce several IBD yield predictions in section~\ref{scn:Prediction}. The response matrix is defined in section~\ref{scn:ResponseMatrix}. Systematic uncertainties are summarized in section~\ref{scn:Systematics}. Finally fitting frameworks are described in section~\ref{scn:Framework}, and results are discussed in section~\ref{scn:results}.


\section{\label{scn:Experiment}Experimental setup}

The \STEREO{} experiment is performed at the high-flux reactor HFR of the Institut Laue Langevin (ILL) \cite{illRTS}. The HFR is designed and operated in order to extract most intense beams of cold, thermal or hot neutrons for research, in particular neutron scattering. It uses a single compact fuel element of 41~cm diameter and 80~cm height with highly enriched \U{}~(enrichment of 93\,\%) \cite{CAMPIONI20091319}. The \STEREO{} detector is installed at approximately 10~m from the core (center-to-center distance). A water transfer channel for reactor operation provides an overburden of 15~m.w.e.\ against cosmic radiation at the \STEREO{} site. The extraction of neutron beams and the operation of neighbouring instruments for neutron scattering impose a high background of fast and thermal neutrons and neutron capture gammas onto the \STEREO{} site, which had to be mitigated by heavy shielding of site and detector. The HFR is operated with alternate periods of about 50 days where the reactor is turned on and periods of reactor turned off.

\begin{figure}[tb]
    \begin{indented}
		\item[] \includegraphics[width=0.75\linewidth]{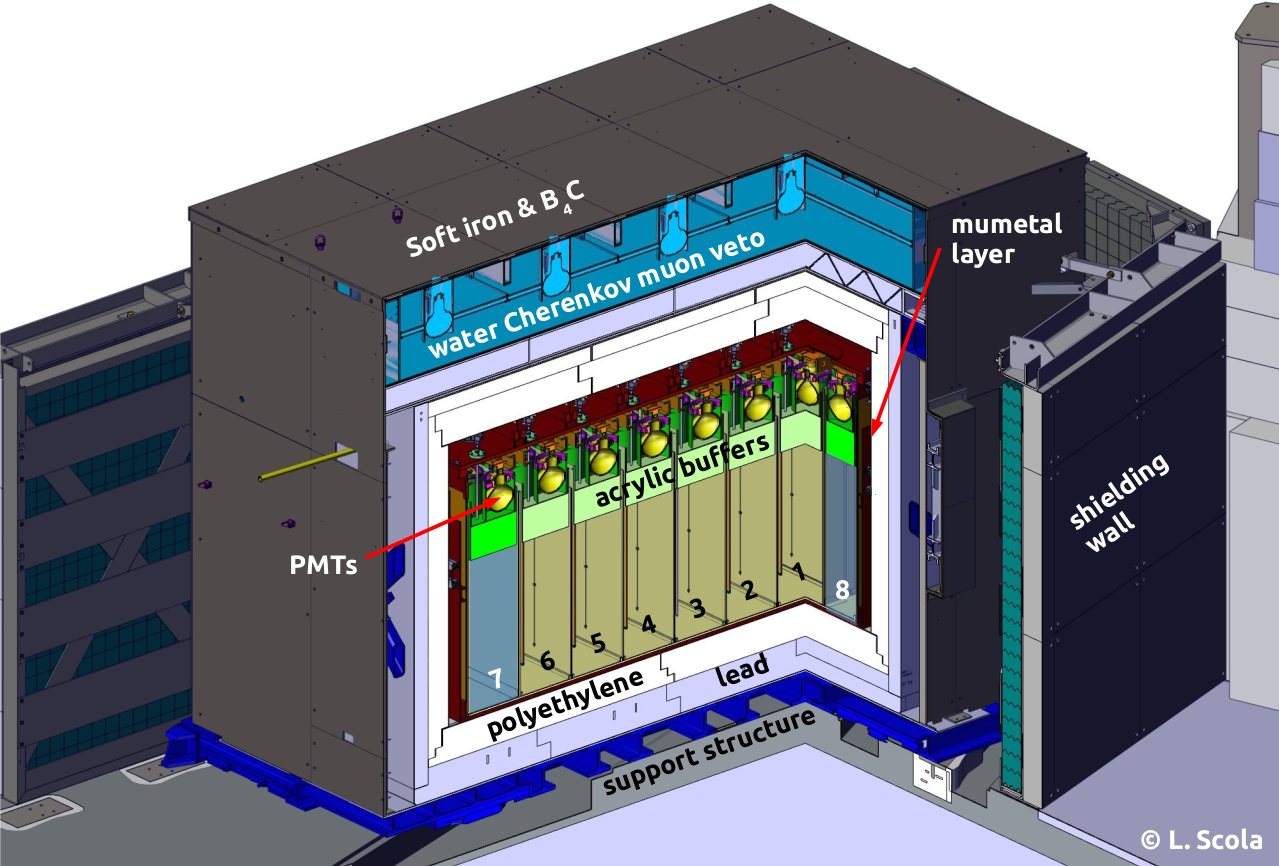}         
		\item[] \includegraphics[width=0.75\linewidth]{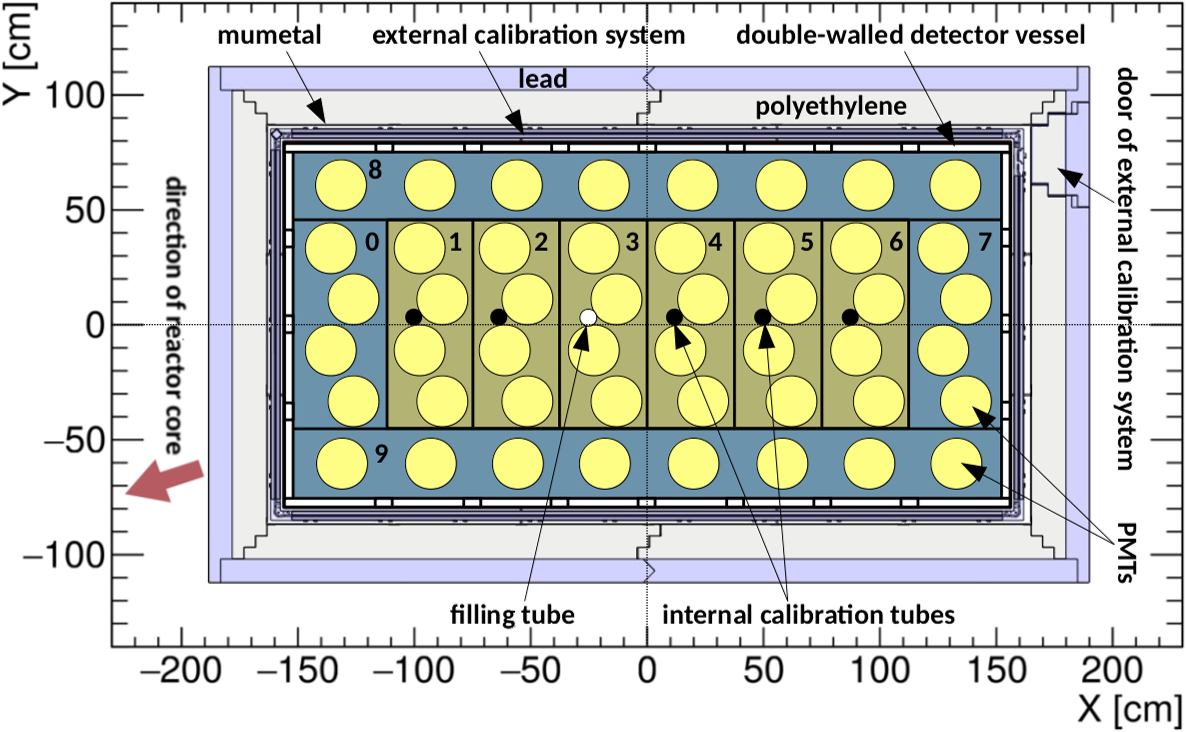}		
    \end{indented}
    \caption{Cutaway view (top) and top view (bottom) of the \STEREO{} detector setup. 1 -- 6: Target cells (baselines from reactor core: 9.4 -- 11.2~m); 0 and 7 -- 9: Gamma-Catcher cells. B\textsubscript{4}C layers on the inner side of shielding walls and the outer detector walls are not shown. Taken from \cite{Almazan_2020}.}
    \label{fig:detector}
\end{figure}

The \STEREO{} detector in only shortly introduced in the following, further details can be found in \cite{Allemandou_2018}. It uses the inverse beta decay (IBD) reaction $\bar{\nu}_{e} + p \rightarrow e^{+} + n$ to detect $\bar{\nu}_{e}$ produced by the HFR. The segmentation of the fiducial volume, called ``Target'' (TG), allows simultaneous spectrum measurements at six different baselines between 9.4 and 11.2~m and is used to search for active-to-sterile neutrino oscillations \cite{Almazan_2020}. In the following analysis, all six cell spectra will be merged together to study the \U-induced spectrum shape, as described in section \ref{scn:Data}. To be able to detect gamma rays from positron annihilation or from the n-Gd cascade that may escape the TG, four other cells are located on an outer crown of the TG volume and are designated as ``Gamma-Catcher'' (GC) (cf.~figure~\ref{fig:detector}). All six TG cells are filled with Gd-loaded liquid scintillator for neutron capture enhancement \cite{Buck_2019} while GC cells are filled with unloaded scintillator. GC cells are also used as a veto against external background. Separation walls between volumes as well as side and bottom walls are built with VM2000\texttrademark~mirror films enclosed in an air-filled gap between two thin acrylic plates (2~mm) to allow high reflectivity and optical separation between active volumes. However, there remains an optical cross-talk of several percent between cells, which is discussed in section \ref{scn:Response}.

The scintillation light is read out by 48~8-inch photomultiplier tubes (PMTs), 4~for each TG cell and short GC cell and 8~for each long GC cell. They are located on top of each cell (cf.~bottom panel of figure~\ref{fig:detector}) and separated from the scintillator by acrylic blocks designated as ``buffers''. The optical coupling between PMTs and acrylic is provided by a bath of mineral oil. PMT signals are continuously digitised at 250~MHz using 14~bit ADCs \cite{Bourrion_2016}. Two levels of gain are available which allows to detect the single photo-electron peak and to have good linearity of the electronics ($<1\,\%$ deviation \cite{Allemandou_2018}) up to 10 MeV. Individual electronic pulses are integrated to determine the amount of light detected by each PMT. The total and the tail pulse integral, $ Q_\mathrm{tot} $ and $ Q_\mathrm{tail} $, respectively, are obtained by integration over the total pulse duration and the pulse tail duration. From $Q_\mathrm{tot}$ we reconstruct the total energy deposit in the cell, and we use $Q_\mathrm{tail}/Q_\mathrm{tot}$ as a metric for particle identification by Pulse Shape Discrimination (PSD), effectively separating interactions from strongly and lightly ionising charged particles (cf. section \ref{scn:Data}).

The detector light response and stability are continuously monitored. A LED-based light injection system \cite{Bourrion_2016} is used to calibrate the PMTs at the photo-electron level and to monitor the linearity of the electronics. In addition, a set of radioactive $\gamma$-ray and neutron sources is regularly deployed inside and all around the detector to monitor the detector response and to determine the energy scale. Sources can be deployed via three different calibration systems: (1) through vertical tubes spanning the full height of the TG, approximately at the center of each TG cell, (2) in a semi-automated positioning system at different heights along the perimeter of the detector, in-between the detector vessel and the shielding, (3) on a rail below the detector along its central long axis (cf.~figure~\ref{fig:detector}).

To reduce external background ($\gamma$-rays and neutrons), the \STEREO{} detector is enclosed in a passive shielding (cf.~figure~\ref{fig:detector}) composed of borated polyethylene and lead. A water Cherenkov detector is placed on top of the lead shielding and used as an active muon veto. Offline cuts using the veto (cf.~section~\ref{scn:Data}) reduce cosmogenic correlated background by about 30\,\% (limited by the geometric coverage of the veto). The remaining cosmogenic background is estimated by comparing data sets from reactor-on and reactor-off periods. A magnetic shielding (soft iron and mu-metal) protects the \STEREO{} system against the effect of the magnetic field generated by other experiments near the \STEREO{} site (cf.~figure~\ref{fig:detector}). The external soft iron layer is covered by boron-loaded rubber (containing B\textsubscript{4}C powder) to absorb the ambient thermal neutrons present in the reactor hall.


\section{\label{scn:Response}Detector response}

\subsection{Tuning of the MC parameters using charge observables}

A detailed description of \STEREO{} Monte Carlo (MC) is available in \cite{Almazan_2020}. MC simulations are built using the Geant4 toolkit \cite{G4} and include all the aspects of an event, from particle interactions in the detector to light collection in the PMTs. The simulation output has the same format as the real data in order to allow for straightforward comparison between them. In the simulation of an event in the detector, the deposited energy is translated into emitted light following a modelling of the scintillation process. Dedicated laboratory measurements of intrinsic properties of the liquid scintillator along with information gathered from calibration sources have been used to fine-tune scintillator parameters, such as the total attenuation length and light yields of both liquid scintillators. Another crucial aspect of the simulation is represented by the optical properties of the separative plates, consisting of VM2000\texttrademark~mirror films and a nylon net, whose purpose is to ensure an air gap, enclosed between two acrylic plates. They have a direct impact on the collection of light, the top-bottom asymmetries and thus on the energy resolution of the detector. In order to model the light cross-talks between the cells, several aspects had to be considered: the angle dependence of the VM2000\texttrademark~films reflectivity, the leakage of liquid scintillator inside some separating walls (partially filling the air gap) and the light absorption inside the separating walls and on the surface of calibration tubes. All these aspects are introduced in the simulation.

The parameters of the optical model were fine-tuned on charge and light leaks distributions for $^{54}$Mn and $^{24}$Na calibration sources deployed at five different positions inside the cells with calibration tubes, such that the simulation reproduces at best the measured charge spectra. It has been shown that the collection of light can be well reproduced both when looking at the cell containing the calibration source and its neighboring cells. The raw-charge agreement between data and MC reaches a \%-level for most of the source positions. The quenching effect for high $\mathrm{d}E/\mathrm{d}x$, which makes the detector response deviate from a perfect linear model is described by an effective Birks coefficient $k_\mathrm{B}$. This coefficient is fine-tuned in the simulation such that the calibration coefficients, \textit{i.e.} the ratio of the mean charge over the mean deposited true energy, for different calibration sources covering a range in energy up to 4.5 MeV, are well reproduced (cf. \cite[Section VI.A]{Almazan_2020}). 

\subsection{Reconstructed energy and energy scale accuracy}\label{scn:EScale}

After this fine-tuning of the MC parameters based on charge observables, the response to the single 835 keV $\gamma$-ray emitted by a $^{54}$Mn source is chosen as the calibration anchor point of the experiment. An iterative procedure adjusts the light leaks and calibration coefficients to make the experimental distributions of reconstructed energies coincide with the simulated distributions. In order to be as representative as possible of the response of a cell as a whole, the adjustment is made after merging the data from 5 positions distributed along the vertical calibration tube.  Potential biases of this energy reconstruction in the \STEREO{} detector would have a direct impact on the shape of the measured antineutrino energy spectrum. A detailed study has already been presented, cell by cell, in \cite{Almazan_2020}. We discuss here the same three main components of the systematic uncertainty at the level of the whole target volume.

\begin{table}[t]
\caption{List of $\gamma$-lines used for the calibration of the \STEREO{} detector across the whole energy range of the antineutrino spectrum. For $^{60}$Co and $^{24}$Na, the energy corresponds to the emission of two coincident gammas. The full deexcitation cascade is considered for neutron captures on Gd (n-Gd). \label{tab:sources}}
\begin{indented}
\item[]\begin{tabular}{rr}
\hline
Source or reaction & Selected $\gamma$-line (MeV) \tabularnewline
\hline \hline 
$^{137}$Cs & 0.662 \tabularnewline
$^{54}$Mn & 0.835 \tabularnewline
$^{65}$Zn & 1.11 \tabularnewline
$^{40}$K & 1.4 \tabularnewline
n-H & 2.2 \tabularnewline
$^{60}$Co & 2.50 \tabularnewline
$^{24}$Na & 4.12 \tabularnewline
Am-Be & 4.43 \tabularnewline
n-Gd & $\approx\,$8 \tabularnewline
\hline \hline 
\end{tabular}
\end{indented}
\end{table}

\begin{figure}[tb]
    \begin{indented}
	\item[] \includegraphics[width=1\linewidth, trim={0 0 0 0.7cm},clip]{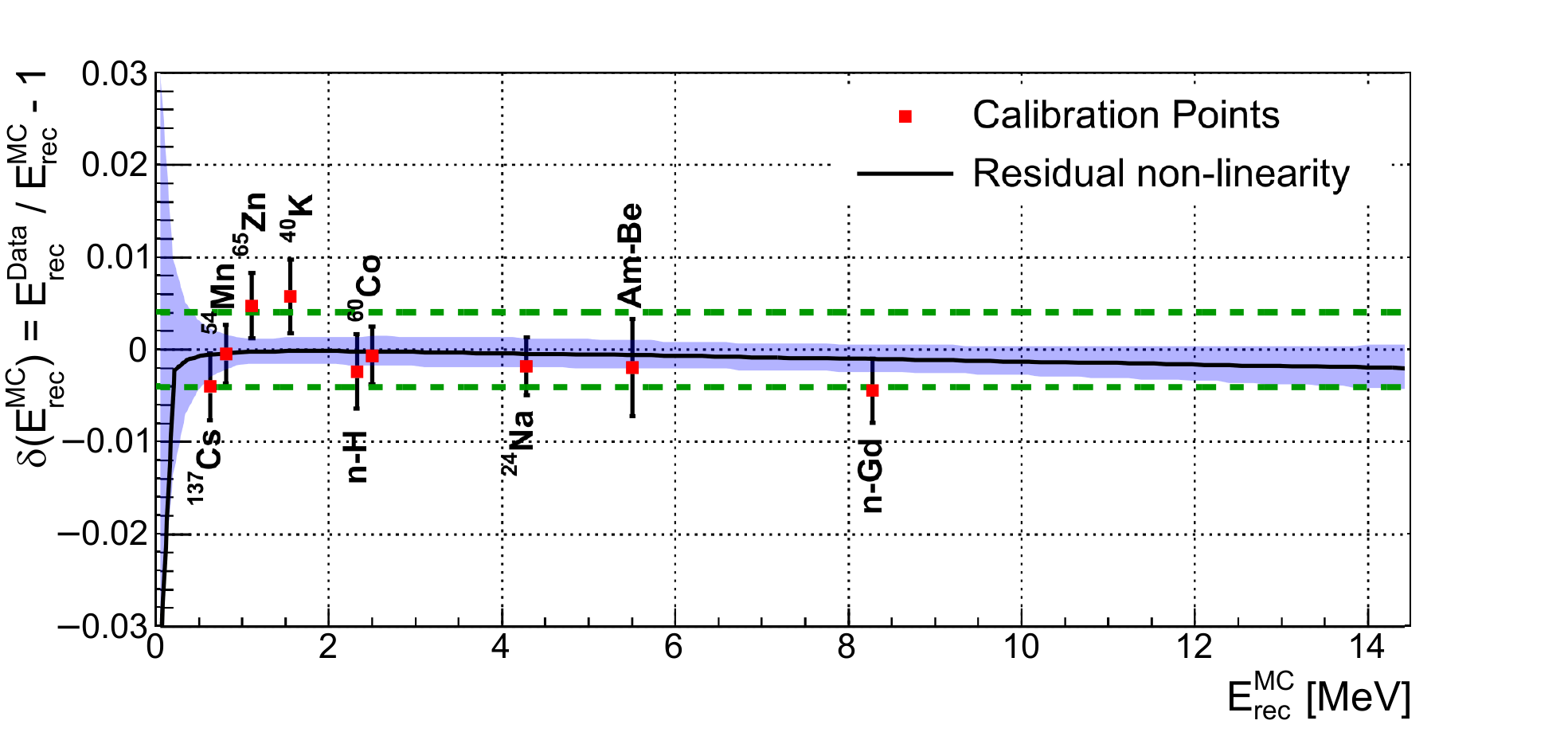}
	\item[] \includegraphics[width=1\linewidth, trim={0 0 0 0.7cm},clip]{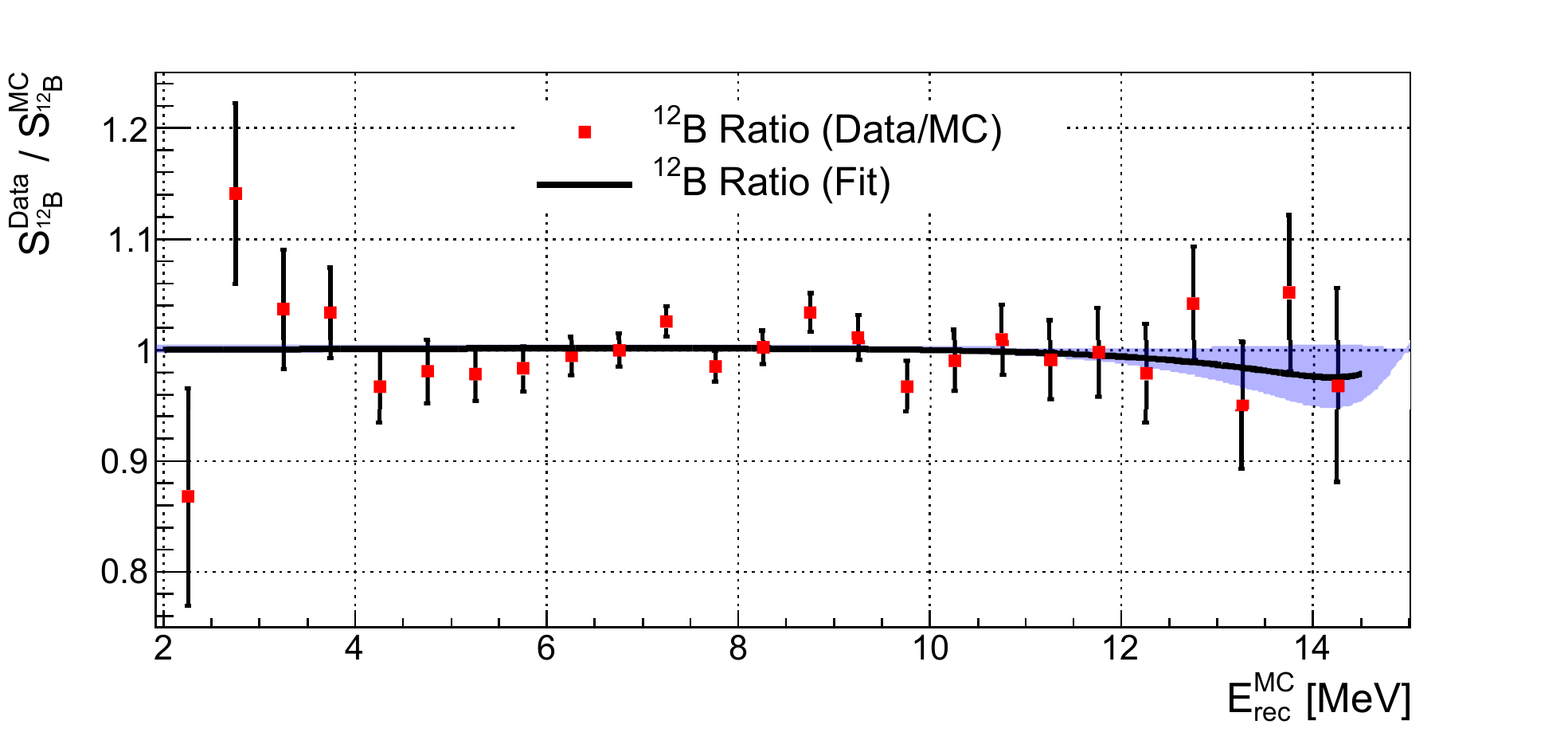}
	\end{indented}
	\caption{Combined fit of all calibration residuals at the target level. The distortion of the energy scale between data and MC is assumed to be a second order polynomial. (Top) Data/MC comparison of reconstructed energies for calibration sources; they differ from the nominal $\gamma$ energy since quenching is included. Fast-neutron-induced proton recoils also contribute to the Am-Be energy. The $^{54}$Mn source, used as anchor of the absolute energy scale, has by construction $E_\mathrm{rec}^\mathrm{Data} / E_\mathrm{rec}^\mathrm{MC}=1$. Green dotted lines illustrate the $1\sigma$ uncertainty band ($\pm 1.02/\sqrt{6}\,\%$) used for the analysis of the spectrum shape. (Bottom) Data/MC ratio of the cosmogenic $^{12}\mathrm{B}$ spectrum.}
	\label{fig:Calib_Residuals}
\end{figure}

The quality of the calibration anchor point is simply determined by looking at the difference between the positions of the reconstructed Mn peak for data and MC. The distribution of this difference for all weekly runs shows an accuracy of 0.2\,\% at the 1$\sigma$ level in each cell. Beyond the $^{54}$Mn anchor point, the accuracy of the energy scale is tested with a set of radioactive sources listed in table \ref{tab:sources}. For each cell and $\gamma$-line the data to MC ratio $E_\mathrm{rec}^\mathrm{Data} / E_\mathrm{rec}^\mathrm{MC}$ of reconstructed peak energies is computed for the average of the 5 vertical deployment positions and for 3 dates of extensive calibration using all sources. Then, the average of all cells is obtained assuming a correlated systematic uncertainty of the fit of the peak position and uncorrelated time stability and statistical uncertainties. These target-averaged ratios, shown on top of figure \ref{fig:Calib_Residuals}, are all contained within 1\,\% from 1, confirming the robustness of the energy reconstruction away from the $^{54}$Mn reference point.

The data-MC agreement is further tested with the continuous spectrum of the $\beta$-decay of $^{12}$B. This isotope is generated in the liquid scintillator via the interaction of cosmic muons with carbon atoms. Selection cuts applied to the \STEREO{} data allow to efficiently select the specific process of muon capture $\mu^- +\, ^{12}\mathrm{C} \rightarrow \, ^{12}\mathrm{B} + \nu_\mu$ in time coincidence with the $^{12}\mathrm{B}$ decay. The muon-capture vertices are probing the whole target volume. Their distribution relies on the characteristics of atmospheric muons and energy loss in the materials of the detector and shielding, which can all be accurately simulated (see \cite[section VI.E]{Almazan_2020} for details). With a $Q_\beta$ of 13.369 MeV the continuous $^{12}$B spectrum spans the full energy range of interest. The ratio of the measured to simulated spectra is shown in the bottom plot of figure \ref{fig:Calib_Residuals}, providing complementary and stringent constraints on the control of the energy scale. A bias of the energy reconstruction in the simulation would propagate as a distortion of the spectrum shape through the formalism described in \cite{Mention:2017dyq}, inducing discrepancies between the experimental and simulated spectra. The deviation of the $^{12}\mathrm{B}$ spectrum ratio from unity is then complementary to the residuals of the calibration peaks as a probe for distortions in the energy scale. A global analysis, merging all the information, is added in figure \ref{fig:Calib_Residuals}. For illustration, the relative distortion between data and MC energy scales are here fitted with a second order polynomial. The uncertainty band of the fit is found comparable with the naive expectation that the initial uncertainty of 1.02\,\% (quadratic sum of 0.2\,\% from Mn anchoring and 1\,\% from the envelope of calibration residuals) determined at the cell level is reduced by a factor $1/\sqrt{6}$ in the absence of cell-to-cell correlation. We checked that this remains true when varying the polynomial order from 2 to 5 and even when using kernel density estimation with a kernel of 8, 15 and 30 gaussian functions, rejecting any significant contribution of high order terms in the data/MC distortions. Therefore, an uncertainty of $\pm 1.02/\sqrt{6}$ \,\% on the calibration coefficients is taken as a safe estimate for the shape analysis of the measured antineutrino spectrum. 

Finally, another feature of the interaction of cosmic-rays in and around the detector, namely the production of spallation neutrons, is used to measure the time stability of the detector response. A fraction of the cosmic neutrons can make their way to the liquid scintillators and get captured on the hydrogen or gadolinium atoms. The relative evolution of the position of the associated 2.2 and 8 MeV peaks is observed over more than one year of detector operation \cite[figure 13]{Almazan_2020}. All variations are contained within 0.3\,\% and fully correlated between cells. The same uncertainty is thus maintained at the target level.

The contributions described above are summed in quadrature and their impact on the uncertainty of the shape of the measured antineutrino spectrum is discussed in section~\ref{scn:Systematics}.


\section{\label{scn:Data}Measured energy spectrum}

The analysis reported in this article uses \STEREO{}'s second phase of data taking, running from mid-2017 to early 2019 (``phase-II'') \cite{ST9,ST10,ST11,ST12}. It amounts to an effective data taking time of 118.5 days with reactor ON and 212.2 days with reactor OFF. Many details on the antineutrino rate extraction method have been given in a previous publication \cite[sections VII, X]{Almazan_2020}, but some are adapted in this paper to treat the target as a whole. Signal identification relies on two steps: first, selection cuts are applied; second, Pulse Shape Discrimination is used to extract the antineutrino IBD spectrum. \\

\textit{Selection cuts.} The first step of the selection relies on the set of cuts listed in table~\ref{tab:selectionCuts}. These cuts are based on the time structure of an IBD event: a prompt signal due to the $e^{+}$ energy deposit and annihilation with an $e^{-}$, then a delayed signal from the neutron capture on a Gd nucleus. Cuts \#1-2 select events in the relevant energy range for the prompt and delayed signals, respectively. Cut \#3-4 select pairs of prompt and delayed signals that have appropriate time and space coincidences. The next set of cuts focuses on the sharing of energy between cells inside the \STEREO{} detector, allowing the 511~keV gammas from IBD-positron annihilation to escape the vertex cell (that has to be a TG cell) and be detected in a neighbour volume (\#5-6), and rejecting events whose main energy deposit is in the GC (\#7). Cuts \#8-9 introduce a veto-time after a muon is detected in the Cherenkov veto or the detector, and cut \#10 requires the prompt-delayed pair to be well isolated in time from any other signal. Finally, cut \#11 rejects low-energy cosmic muons with asymmetric light collection due to energy deposit only at the top of a cell. The overall selection efficiency for antineutrinos in the 2-8~MeV range interacting in TG cells is $55.8\pm 0.3\,\%$. The unfolding of the measured prompt spectrum to the antineutrino energy spectrum takes into account the energy dependence of this efficiency via the response matrix (see section \ref{scn:ResponseMatrix}). Systematic uncertainties induced by possible mis-computation of cut efficiencies are discussed in the section~\ref{scn:Systematics}.

\begin{table}[t]
    \caption{Selection cuts for IBD candidates.}
    \label{tab:selectionCuts}
    \begin{indented}
    \item[] 
    \begin{tabular}{l|r|l}
    \hline
    Type        &\#& Requirement for passing cut \\
    \hline
    Energy      &1& $1.625~\mathrm{MeV} < E^\mathrm{detector}_\mathrm{prompt} < 7.125~\mathrm{MeV}$ \\
                &2& $4.5~\mathrm{MeV} < E^\mathrm{detector}_\mathrm{delayed} < 10.0~\mathrm{MeV}$\\
    \hline
    Coin-       &3& $2~\mu\mathrm{s} < \Delta T_\mathrm{prompt-delayed} < 70~\mu\mathrm{s}$ \\
    cidence     &4& $\Delta X_\mathrm{prompt-delayed} < 600~\mathrm{mm}$ \\
    \hline
    Topology    &5& \multirow{2}{*}{$E^\mathrm{cell}_\mathrm{prompt} < \left\{\begin{array}{l}
        1.0~\mathrm{MeV},\mathrm{cell~neighbour~to~vertex~cell}\\
        0.4~\mathrm{MeV},\mathrm{cell~far~from~vertex~cell}
        \end{array}\right.$} \\
                &6& \\
                &7& $E^\mathrm{Target}_\mathrm{delayed} > 1.0~\mathrm{MeV}$ \\
    \hline
    Rejection   &8& $\Delta T^\mathrm{veto}_\mathrm{muon-prompt} > 100~\mu\mathrm{s}$ \\
    of muon-    &9& $\Delta T^\mathrm{detector}_\mathrm{muon-prompt} > 200~\mu\mathrm{s}$ \\
    induced     &10& $\Delta T_\mathrm{before~prompt}$ $> 100~\mu\mathrm{s}$ and $\Delta T_\mathrm{after~delayed}$ $> 100~\mu\mathrm{s}$ \\
    background        & & for all events with $E^\mathrm{detector}_\mathrm{event}$ > 1.5 MeV  \\
      
                &11& $\frac{Q_\mathrm{PMT max, prompt}}{Q_\mathrm{cell, prompt}} < 0.5$\\
    \hline
    \end{tabular}
    \end{indented}
\end{table}

Events selected by the cuts listed above break down into antineutrino events and background events, which can be accidental (\textit{i.e.} random) pairs or physically correlated pairs. The rate of accidental pairs is measured along the data flow using an off-time method \cite[section X.A]{Almazan:2018wln}. Correlated background pairs are characterized by reactor-off event distributions. The presence of a potential reactor-related correlated background cannot be evaluated this way and is discussed below. For now, no such background is assumed.\\

\begin{figure}[tb]
	\begin{indented}
	\item[] \includegraphics[width=\linewidth]{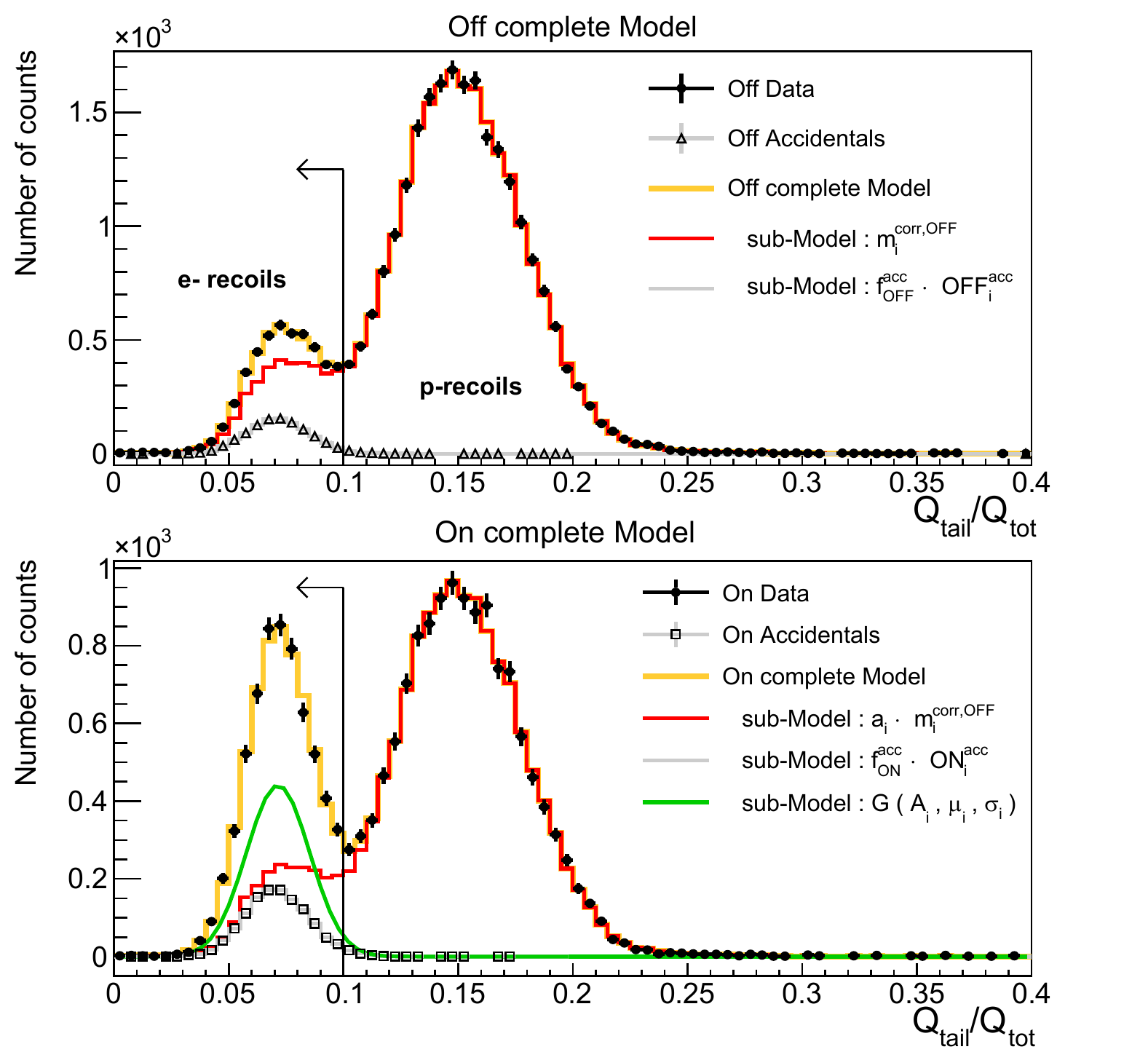}
	\end{indented}
	\caption{Illustration of the simultaneous fit to extract the IBD rate for the bin centered at 3.0~MeV. (Top) Reactor-off PSD distribution, containing accidentals (grey) and correlated pairs (red). (Bottom) Reactor-on PSD distribution broken down into correlated background (red, scaled from reactor-off), accidentals (grey), and the antineutrino component (green). Vertical lines indicate the electron recoils region, corresponding to $2\sigma$ above the mean of the electron recoils peak.}
	\label{fig:PSD}
\end{figure}

\textit{Pulse Shape Discrimination.} The second step of the analysis relies on the Pulse Shape Discrimination (PSD) on the prompt signal. It allows to discriminate between electron-like recoils, where IBD events are located, and proton-like recoils (cf. top panel of figure~\ref{fig:PSD}).

The PSD variable $Q_\mathrm{tail}/Q_\mathrm{tot}$ evolves over time, mainly due to temperature variations of the liquid scintillator. The evolution is daily monitored with uncorrelated $\gamma$~events and correspond to a global shift (the peak position varies by about $ -0.1\, \sigma_\gamma/^\circ \mathrm{C}$). The PSD value of each event is corrected. Moreover, because of light leaks, the average PSD value is different for each cell. Cell 1 is taken as reference and the PSD value of each event in cell $\ell$ is further corrected by the cell average difference $\langle \mathrm{PSD} \rangle_\ell - \langle \mathrm{PSD} \rangle_1$. These corrections allow to merge events from all runs and cells to build the PSD distribution of the full target. To extract the antineutrino component, a joint fit of reactor-on and reactor-off PSD distributions is performed. PSD distributions of IBD candidates and accidental coincidences measured with reactor-off (reactor-on) in energy bin $i$ are denoted as $\mathrm{OFF}_{i,p}$ ($\mathrm{ON}_{i,p}$) and $\mathrm{OFF}^\mathrm{acc}_{i,p}$ ($\mathrm{ON}^\mathrm{acc}_{i,p}$), respectively, with $p$ the PSD bin index. The PSD distribution in reactor-on contains (cf. figure~\ref{fig:PSD}): \begin{itemize}
    \item[-] the antineutrino signal, modelled by a gaussian distribution with integral $A_i$, mean $\mu_{i}$ and standard deviation $\sigma_{i}$;
    \item[-] accidental background, directly evaluated from $\mathrm{ON}^\mathrm{acc}_{i,p}$;
    \item[-] correlated background, modelled by a rescaling of reactor-off PSD distribution $m^\mathrm{corr,OFF}_{i,p}$ which is determined from $OFF_{i,p}$, accounting for its accidental component $\mathrm{OFF}^\mathrm{acc}_{i,p}$. The scale parameter $a_i$ accounts for example for variations of cosmic muon rates due to atmospheric pressure.
\end{itemize} 

\noindent The following fit is thus performed: \begin{equation}\label{eqn:PSDfit-ON}
      \mathrm{ON}_{i,p} = a_{i} \, m^\mathrm{corr,OFF}_{i,p} + f^\mathrm{acc}_\mathrm{ON} \, \mathrm{ON}^\mathrm{acc}_{i,p}  + G_\nu(A_{i},\mu_{i},\sigma^{2}_{i})
\end{equation}
\begin{equation}\label{eqn:PSDfit-OFF}\mathrm{OFF}_{i,p} = m^\mathrm{corr,OFF}_{i,p} + f^\mathrm{acc}_\mathrm{OFF} \, \mathrm{OFF}^\mathrm{acc}_{i,p} 
\end{equation}
\noindent where $f^\mathrm{acc}_\mathrm{ON}$ and $f^\mathrm{acc}_\mathrm{OFF}$ are measured scaling factors of accidental distributions (both $\sim 1/10$). Free fit parameters are $a_{i}$, $A_{i}$, $\mu_{i}$, $\sigma_{i}$ and all $m_{i,p}^\mathrm{corr,OFF}$. Only PSD bins $p$ where $\mathrm{OFF}_{i,p}$ and $\mathrm{ON}_{i,p}$ are nonzero are used. The rate of IBD events $A_{i}$ is extracted for each energy bin $i$ separately. This joint fit allows in particular to simultaneously fit the scaling parameter $a_i$ and the antineutrino component. In particular, it removes systematic effects related to the atmospheric pressure correction between ON and OFF periods.

The analysis at the target level, merging PSD distributions from individual cells before performing the fit, allows to use finer bins than previous analyses (250~keV instead of 500~keV in \cite{Almazan:2018wln,Almazan_2020}). Indeed, binned log-likelihood maximizations are known to introduce biases in the low-statistics regime \cite{FisherMaximumLikelihood}. Reproducing the fitting procedure with the 250~keV binning in simulations, we found biases to be small ($<1\,\%$) but use them even so to correct antineutrino rates extracted from the fit.

\begin{figure}[tb]
	\begin{indented}
	\item[]	\includegraphics[width=0.8\linewidth]{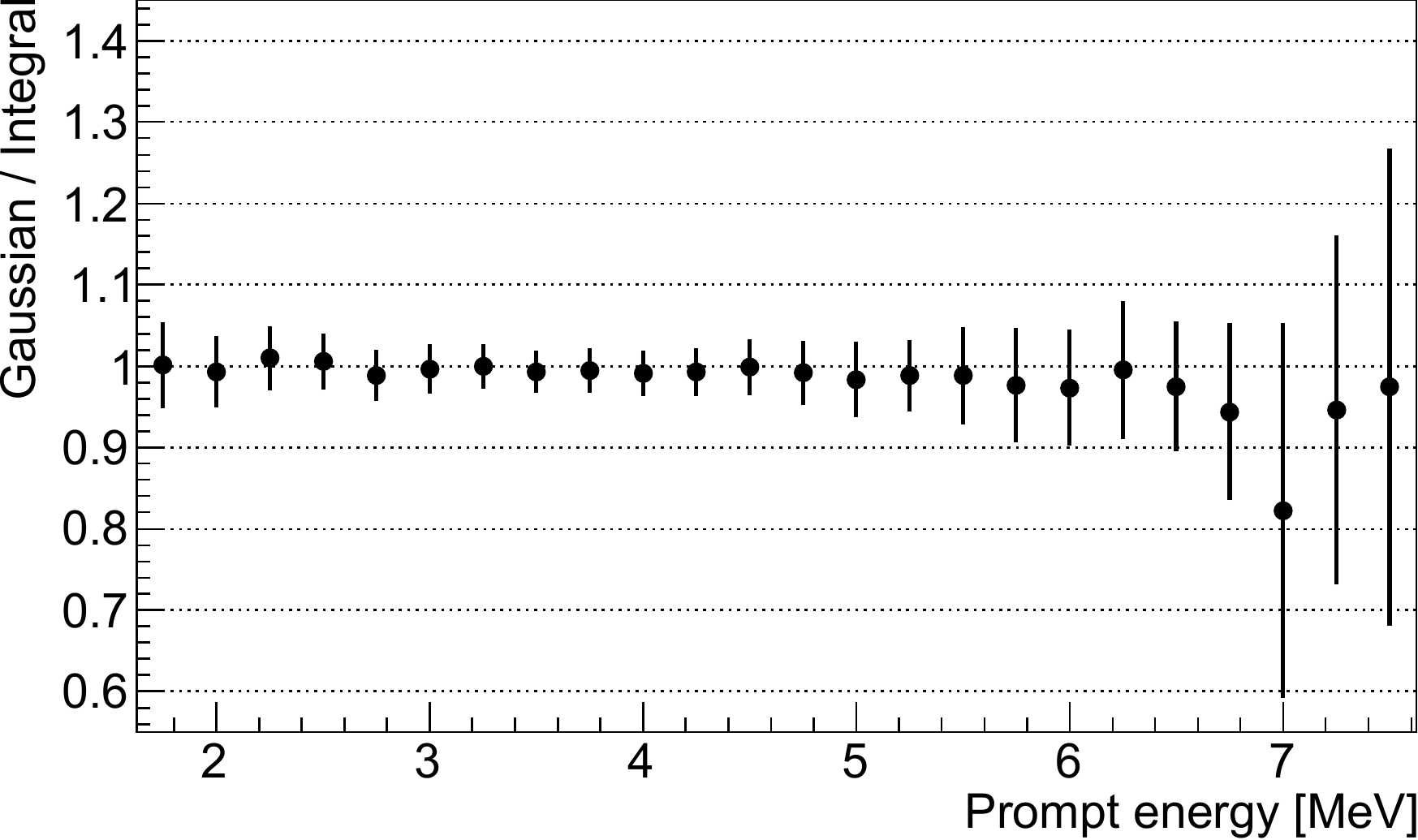}
	\end{indented}
	\caption{Ratio of antineutrino rates from the PSD fit assuming gaussian antineutrino PSD ('gaussian') and from $\mathrm{ON}_i - a'_i\,\mathrm{OFF}_i$ histogram integration ('integral'). These two quantities being highly correlated, the deviations from unity are expected to be small with respect to the displayed errors bars, taken as the relative uncertainty on the $A_i$ parameter evaluated by the fit. Except for the bin at 7~MeV (see discussion in the text), a potential systematic bias is negligible compared to this statistical uncertainty.}
	\label{fig:gaussian-nu-PSD}
\end{figure}

The validity of the gaussian model for the antineutrino PSD has also been investigated. The antineutrino rate extracted from the fit ($A_i$ parameter) has been compared to the sum of the bin content of the $\mathrm{ON}_i - a'_i\,\mathrm{OFF}_i$ histograms in the electron recoils region, accidental pairs being subtracted beforehand. Since the tail of the gaussian model is spreading beyond the electron recoils region, antineutrino rates $A_i$ have been corrected to match the range of the histogram summation. Since scaling parameters $a_i$ from the fit are correlated to the assumption of gaussian antineutrino PSD (cf. equations (\ref{eqn:PSDfit-ON})-(\ref{eqn:PSDfit-OFF})), new scaling parameters $a'_i$ have been computed based on the proton recoils region only. They are consistent with the nominal $a_i$'s at the 0.5\% level. The result of the comparison of antineutrino rates is presented in figure~\ref{fig:gaussian-nu-PSD} and shows a very good agreement across the energy range. The most significant discrepancy is found in the 7 MeV bin. This effect is most likely related to a strong statistical fluctuation distorting the shape of the gaussian fit of the antineutrino PSD distribution in this bin. Indeed, we observed that (1) relative uncertainties on fit parameters $(A_i, \mu_i, \sigma_i)$ in this bin are about twice as large as in the previous bin, indicating that the fit may be unstable; (2) the best-fit width $\sigma_i$ is lower than in bins above and below; (3) the difference between the two methods is driven by the last PSD bin in the electron recoils region, not taken into account by the gaussian model whose best-fit width $\sigma_i$ may be too small; (4) a satisfactory agreement is recovered for higher energy bins, ruling out the option of a significant systematic effect related, for instance, to low signal-to-background ratios. Overall, a satisfactory agreement is found and validates the use of the gaussian model.\\

\begin{figure}[tb]
    \begin{indented}
		\item[] \includegraphics[width=0.65\linewidth]{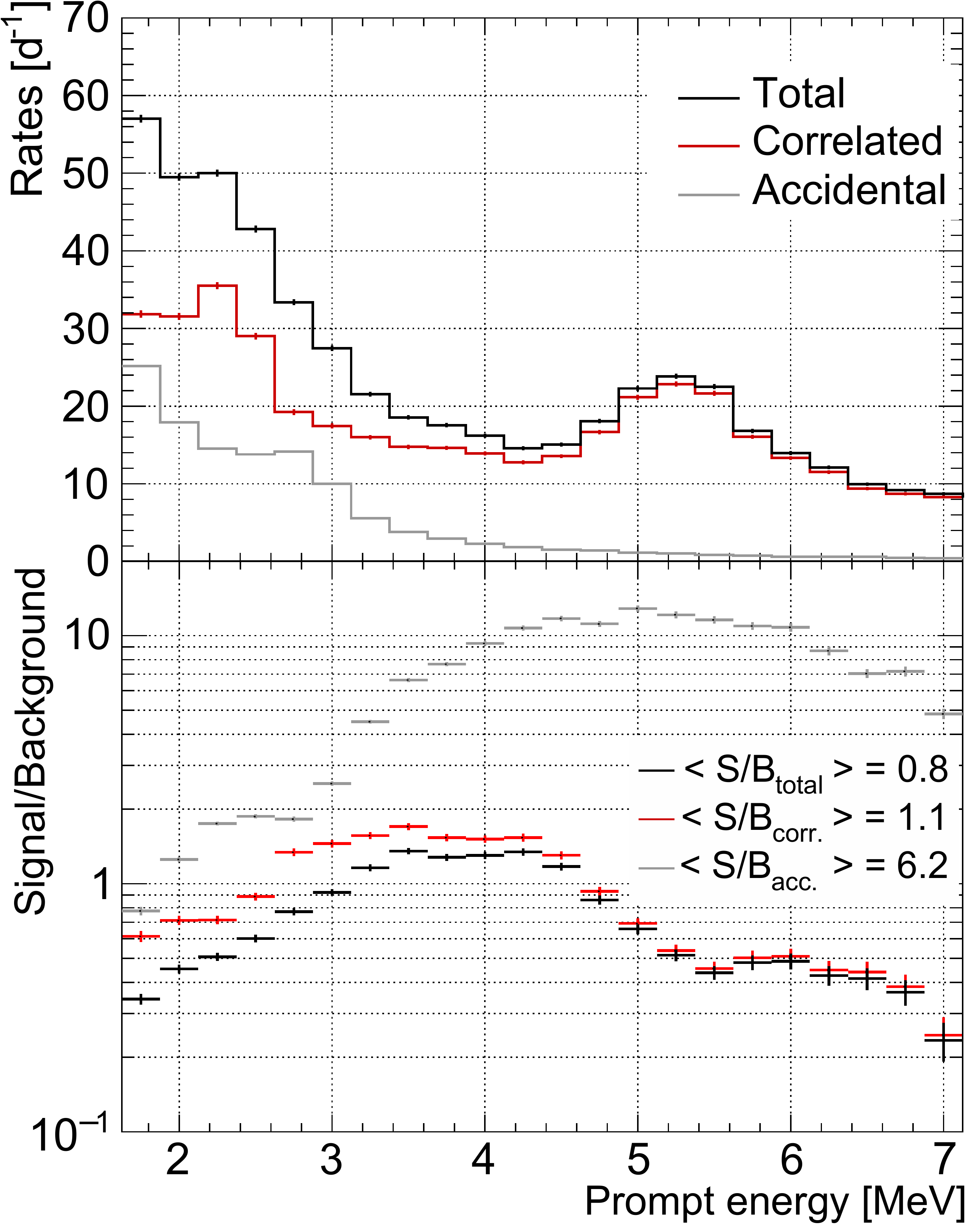}
    \end{indented}
    \caption{(Top) Background spectrum of correlated and accidental contributions, defined as the rate integral of the electronic recoil component. (Bottom) Measured signal-to-background ratio with respect to the background spectra shown above. The average values given in the legend are calculated from the data points weighted by the signal. Adapted from \cite{Almazan_2020}.}
    \label{fig:SBR}
\end{figure}

\textit{Backgrounds.} Background distributions for correlated and accidental pairs are obtained from the red ($a_{i} \, m^\mathrm{corr,OFF}_{i,p}$) and grey ($f^\mathrm{acc}_\mathrm{ON}\, \mathrm{ON}^\mathrm{acc}_{i,p}$) distributions in the bottom panel of figure~\ref{fig:PSD}, respectively. In order to evaluate signal-to-background ratios, we use as antineutrino rate (for this study only) the sum of bin contents until the upper limit of the electron recoils range shown in figure~\ref{fig:PSD}. Doing so, we can use the same PSD range of integration for signal and backgrounds. The measured background and signal-to-background ratios are displayed in figure \ref{fig:SBR}. We require (arbitrarily) a signal-to-background ratio of $>0.2$, leading to an energy range of [1.625~MeV, 7.125~MeV] in prompt energy, \textit{i.e.} 22 bins of 250~keV. 

The accidental energy spectrum is dominated by low-energy gamma events. Events above the highest natural radioactivity $\gamma$-line at 2.6~MeV (from Tl) are due to two main sources: (1) cascades following neutron captures on Gd, or (2) high-energy $\gamma$-rays from neutron capture on surrounding materials (Al, Fe) during reactor-on periods, due to neighbouring instruments.

The high-energy background spectrum is dominated by correlated events. A first peak of correlated background at 2.2~MeV arises from muon-induced multi neutron captures in the liquid surviving the isolation cut (\#10 in table~\ref{tab:selectionCuts}). A second large peak around 5.4~MeV is due to $^\mathrm{12}\mathrm{C}(\mathrm{n},\mathrm{n'}\gamma)^\mathrm{12}\mathrm{C}$ interactions in the liquid, with the prompt energy coming from the coincident $\gamma$-ray and fast-neutron-induced proton recoils.

As described above, accidental and cosmogenic correlated backgrounds are accurately taken into account via the joint fit of reactor-on and reactor-off PSD distributions. As for reactor-related background, it cannot be evaluated using reactor-off events. It is however possible to compare reactor-on and reactor-off PSD distributions in the proton recoils region. The study presented in \cite{Almazan_2020} shows a small excess of events when reactor is on, which amounts to about 3\,\% in the lowest energy bin and decreases with energy as a power law. Antineutrino rates are corrected for this according to signal-to-background ratios. However, the relevant background for antineutrino search is located in the electron recoils region (where IBD events also lie), and an observed excess in proton recoils does not necessarily imply the same for electron recoils. To cover all scenarios we add a 100\,\% uncertainty on this correction, which is displayed in section~\ref{scn:Systematics}.


\section{\label{scn:Prediction}Predicted antineutrino energy spectrum}

Several predictions exist for the most relevant fissioning isotopes, $\mathrm{X} \in\{^{235}{\rm U},^{238}{\rm U},^{239}{\rm Pu},^{241}{\rm Pu}\}$, based on the conversion method or the summation method (also a combination of both has been used). The reference antineutrino spectra used today in reactor antineutrino experiments come from the conversion method (Huber-Mueller model, HM) \cite{Huber:2011wv,Mueller:2011nm}. Recent dedicated measurements of the $\beta$-strength of important fission products have improved nuclear data for the summation method substantially, calling for a test of its latest predictions \cite{Estienne:2019ujo} (SM) against the HM model. Furthermore, experimental antineutrino spectra associated to the fission of \U~and $^{239}$Pu have been provided by the Daya Bay (DB) collaboration \cite{Adey:2019ywk}. However, they have not been unfolded. We thus introduce our own DB-like bumped models for this analysis (cf. table~\ref{tab:models}). The antineutrino spectrum obtained by method $M\in\{{\rm HM},{\rm SM}\}$ for isotope X is denoted with $S_{M}^\mathrm{X}(E_{\nu})$.

Several HFR-specific corrections have to be applied before comparing with the antineutrino spectrum measured by {\sc Stereo}. Firstly, the predicted HM and SM spectra provide a snapshot of the fission spectrum after only 12 hours of irradiation of the target X used in the measurements of the fission beta spectra \cite{Schreckenbach:1985ep,Hahn:1989zr}. Therefore, they have to be corrected for the accumulation of fission products with lifetime comparable to or larger than 12 hours. The relative correction of these off-equilibrium effects is called $\delta(E_\nu)$. The mean evolution of the fuel over one reactor cycle was calculated using the \texttt{FISPACT-II} code~\cite{FISPACT} considering a constant power during all the cycle. The results of this calculation coupled to the \texttt{BESTIOLE} code~\cite{BESTIOLE}, which calculates the antineutrino energy spectrum for each isotope, allow us to determine the relative distortion $\delta(E_{\nu})$. We note that neutron capture by a fission product suppresses the decay of this fission product and replaces it by another likely beta-active isotope. Such effects yield a flux-dependent correction, thus depending on the operation power of the reactor. A class of these so-called non-linear nucleides have been investigated in~\cite{Huber:2016} and been found to be negligible for the HFR. The off-equilibrium correction is found to mainly affect energy bins below $E_\nu =3.6~\mathrm{MeV}$, with a contribution of 1.8\,\% at 2 MeV.

Secondly, the fuel contains initially 7\,\%\ of $^{238}$U. Breeding of $^{239}$Pu by neutron capture results in a relative contribution $p_{\rm Pu9} = 0.007$ of $^{239}$Pu to all fissions, averaged over one reactor cycle. It yields to a 0.3~\% deficit with respect to a pure-\U{}~$\bar{\nu}_e$ production, with little impact on the spectrum shape ($<0.15~\%$). As this article reports on the shape of the IBD spectrum, contribution from $^{239}$Pu (and other fissionning isotopes) can be neglected.

Thirdly, beta decays from reactor structural materials activated by neutron capture contribute significantly to the measured spectrum. This contribution $S_{\rm A}(E_\nu)$ was evaluated using a complete 3D \texttt{TRIPOLI-4\textregistered}~\cite{TRIPOLI} simulation of the reactor core. The reactor core and most of the beam tubes being made of aluminum, $^{28}$Al decay is the main contributor. There is also a sub-leading contribution from activation of $^{55}$Mn. Overall, these activated elements produce antineutrinos with energies as high as 2.86~MeV, leading to a mean correction of about 10\,\% below this energy. 

Fourthly, spent fuel is stored in the transfer channel above the \STEREO{} detector. The antineutrino spectrum $S_{\rm SF}(E_\nu)$ from the decay of its beta-active isotopes has to be taken into account. The spent fuel term was evaluated in the same way as off-equilibrium contribution, using \texttt{FISPACT-II} coupled to \texttt{BESTIOLE}. Its contribution was estimated to be less than 0.1\,\% after 24h of a reactor stop, justifying that in our analysis only data after this time are considered.

Finally, the total spectrum predicted by model $M$ and taking into account the aforementioned corrections, averaged over one reactor cycle of the HFR, writes:
\begin{equation}\label{eqn:flux-model}
    S_M(E_{\nu}) = S_M^\mathrm{U5}(E_{\nu}) + S^\mathrm{corr}(E_\nu)
\end{equation}
with the correction term (linear terms only) being \begin{eqnarray}
  S^\mathrm{corr}(E_\nu) &= p_{\rm Pu9}\left[ S_{M}^{\rm U5}(E_{\nu}) - S_M^{\rm Pu9}(E_{\nu})\right] + \delta(E_\nu)\, S_M^\mathrm{U5}(E_{\nu}) \nonumber \\ & \quad + S_{\rm A}(E_{\nu}) + S_{\rm SF}(E_{\nu}).\label{eqn:corr-full}
\end{eqnarray}
This term simplifies to \begin{equation}
\label{eqn:corr-reduced}
  S^\mathrm{corr}(E_\nu) = \delta(E_\nu)\, S_{\rm HM}^\mathrm{U5}(E_{\nu})
          + S_{\rm A}(E_{\nu}) 
\end{equation} when neglecting $^{239}$Pu and spent fuel contributions; at first order, the off-equilibrium component is taken to be the same for all models. Further details on the calculations of these corrections can be found in~\cite{Almazan_2020}. \\

The IBD cross section $\sigma_\mathrm{IBD}$ from \cite{Strumia_2003} is used to obtain the total IBD prediction $\Phi^\mathrm{tot}_M = S_M \times \sigma_\mathrm{IBD}$ from a given antineutrino energy model $M$. It naturally breaks down into $\Phi^\mathrm{tot}_M = \Phi^\mathrm{U5}_M + \Phi^\mathrm{corr}$ following equation~(\ref{eqn:flux-model}). Table \ref{tab:models} defines the models used in this spectral analysis as inputs to build or validate analysis frameworks (cf. sections \ref{scn:ResponseMatrix}, \ref{scn:Framework}). 

\begin{table}[tb]
    \caption{IBD yield models $\Phi^\mathrm{U5}_M$ used in this analysis. Specific bumped models (\texttt{HMBump9}, \texttt{SMBump9}) are introduced to reproduce a Daya Bay-like excess.}
    \label{tab:models}
    \begin{indented}
    \item[] \begin{tabular}{c|c|c}
    \hline \hline
    Name        & Flux model $S_{M,{\rm U5}}(E_{\nu})$ & $\sigma_\mathrm{IBD}$ \\
    \hline
    \texttt{HM} & HM \cite{Huber:2011wv} & \multirow{4}{*}{Strumia-Vissani \cite{Strumia_2003}} \\
    \texttt{SM} & SM \cite{Estienne:2019ujo} &  \\
    \texttt{HMBump9} & HM + 9\,\% bump at $6\pm 0.3$ MeV &   \\
    \texttt{SMBump9} & SM + 9\,\% bump at $6\pm 0.3$ MeV &   \\
    \hline \hline
    \end{tabular}
    \end{indented}
\end{table}


\section{Response matrix} \label{scn:ResponseMatrix}

In order to extract the antineutrino energy spectrum with respect to the incident antineutrino energy $E_\nu$, an essential ingredient is the detector response matrix. The prompt signal energy $E_\mathrm{pr}$ is related to $E_\nu$ as \begin{equation}\label{eqn:Epr_vs_Enu}
E_\mathrm{pr} \simeq E_{\nu}-{\Delta}M \cdot c^2+m_{e}  \cdot c^2=E_{\nu}-0.782~\mathrm{MeV} \end{equation} 
where ${\Delta}M = m_n - m_p$ (the kinetic energy of the IBD neutron is negligible). Because of several unavoidable detector effects (quenching, resolution, energy loss, inefficiency) the relation (\ref{eqn:Epr_vs_Enu}) cannot be used directly to unfold the prompt spectrum event by event. Instead, the response matrix $R$ (hereafter \textit{response}) contains the transition probability for an antineutrino with energy $E_\nu$ to be detected with prompt energy $E_\mathrm{pr}$:\begin{equation}
    R_{ij} = \mathrm{P} \big(E_\mathrm{pr} \mathrm{~in~bin~} j \, | \,  E_\nu \mathrm{~in~bin~} i\big).
\end{equation} 
The selection efficiency $e_i$ in bin $i$ is smaller than 1 since there is a chance that an interacting antineutrino with given $E_\nu$ is rejected by selection cuts. $R$ is thus normalized such that $\sum_j R_{ij} = e_i$. This choice of normalization allows to encapsulate all detector effects in the same object, the response matrix.

\begin{figure}[tb]
    \begin{indented}
		\item[] \includegraphics[width=0.9\linewidth]{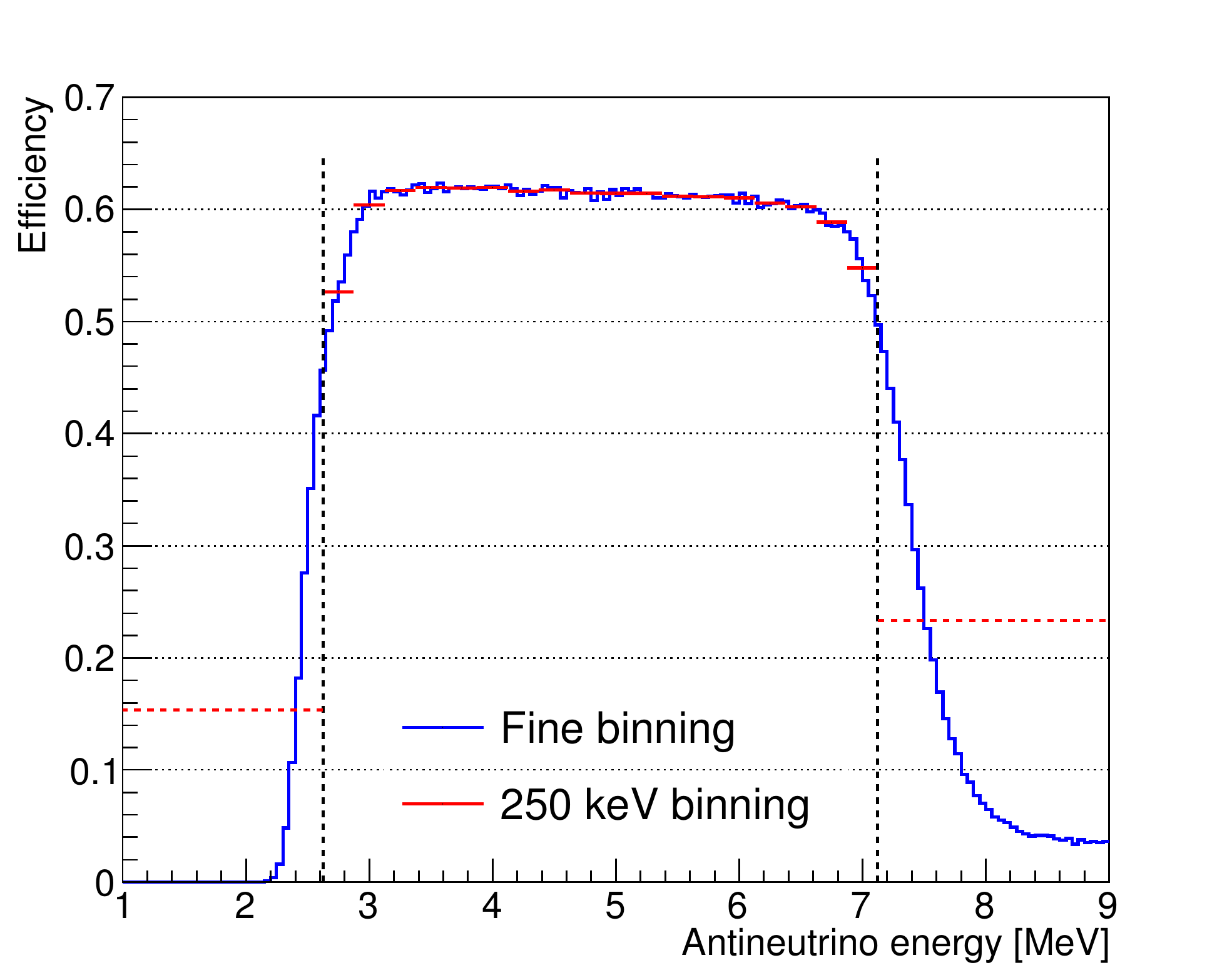}
    \end{indented}
    \caption{Selection efficiency $e_i$ as a function of the antineutrino energy $E_\nu$ (blue: fine binning; red: 250 keV wide binning). The analysis range [2.625 MeV, 7.125 MeV] is defined by vertical dotted lines: it encloses region of highest efficiency. The larger edge bins are used for regularization purposes only.}
    \label{fig:efficiency}
\end{figure}

The response is sampled using \STEREO{}'s detector simulation. Interacting antineutrinos are generated with a flat energy distribution (from the IBD threshold to 10~MeV) and weighted according to \STEREO{}'s nominal IBD yield prediction $\Phi^{\mathrm{tot},0} \equiv \Phi_\mathrm{HM}^{\mathrm{tot}}$ based on the Huber-Mueller model. Interaction products are propagated through the detector and selection cuts \#1-7 from table \ref{tab:selectionCuts} are applied. IBD candidates passing through the selection are stored in the response matrix, every column of which is finally normalized to $e_i$. The efficiency $e_i$ in true bin $i$ is computed, using the same simulation, as the ratio of selected IBD candidates over antineutrinos interacting in the TG. 
The response relates any IBD spectrum $\Phi_i^{\rm tot}$ to the corresponding predicted prompt spectrum $N_j$ as \begin{equation}
    N_j = \sum_i R_{ij} \Phi_i^{\rm tot}.
\end{equation}

The \textit{analysis range} in antineutrino energy is defined as [2.625 MeV, 7.125 MeV] and corresponds to the region of highest selection efficiency ($>50\,\%$), as illustrated in figure~\ref{fig:efficiency}. The lower bound is shifted up with respect to the range of the measured energy spectrum [1.625 MeV, 7.125 MeV]: this is due to the energy shift from equation~(\ref{eqn:Epr_vs_Enu}). The $E_\nu$ range is divided in 18 bins of 250 keV. Low-energy and high-energy bins are added, integrating from 1.806 MeV to 2.625 MeV and 7.125 MeV to 10 MeV respectively: they are used for regularization purposes only (cf. section~\ref{scn:Framework}). The response is then a $22 \,  (E_\mathrm{pr} \, \mathrm{ bins}) \times  \, 20 \, (E_\nu\, \mathrm{ bins})$ matrix.

\section{Systematic uncertainties} \label{scn:Systematics}
A first set of systematic uncertainties relates to the extraction of antineutrino rates.

\textit{Reactor-related background.} As introduced in section~\ref{scn:Data}, antineutrino rates are corrected for possible reactor-related background, that cannot be modelled by reactor-off event distributions. To cover all scenarios, a 100\,\% uncertainty on this correction is applied, uncorrelated between energy bins. It results in about 4.5\,\% uncertainty in the first prompt energy bin, decreasing with energy as a power law \cite{Almazan_2020}.

\begin{figure}[tb]
    \begin{indented}
    \item[]	\includegraphics[width=\linewidth]{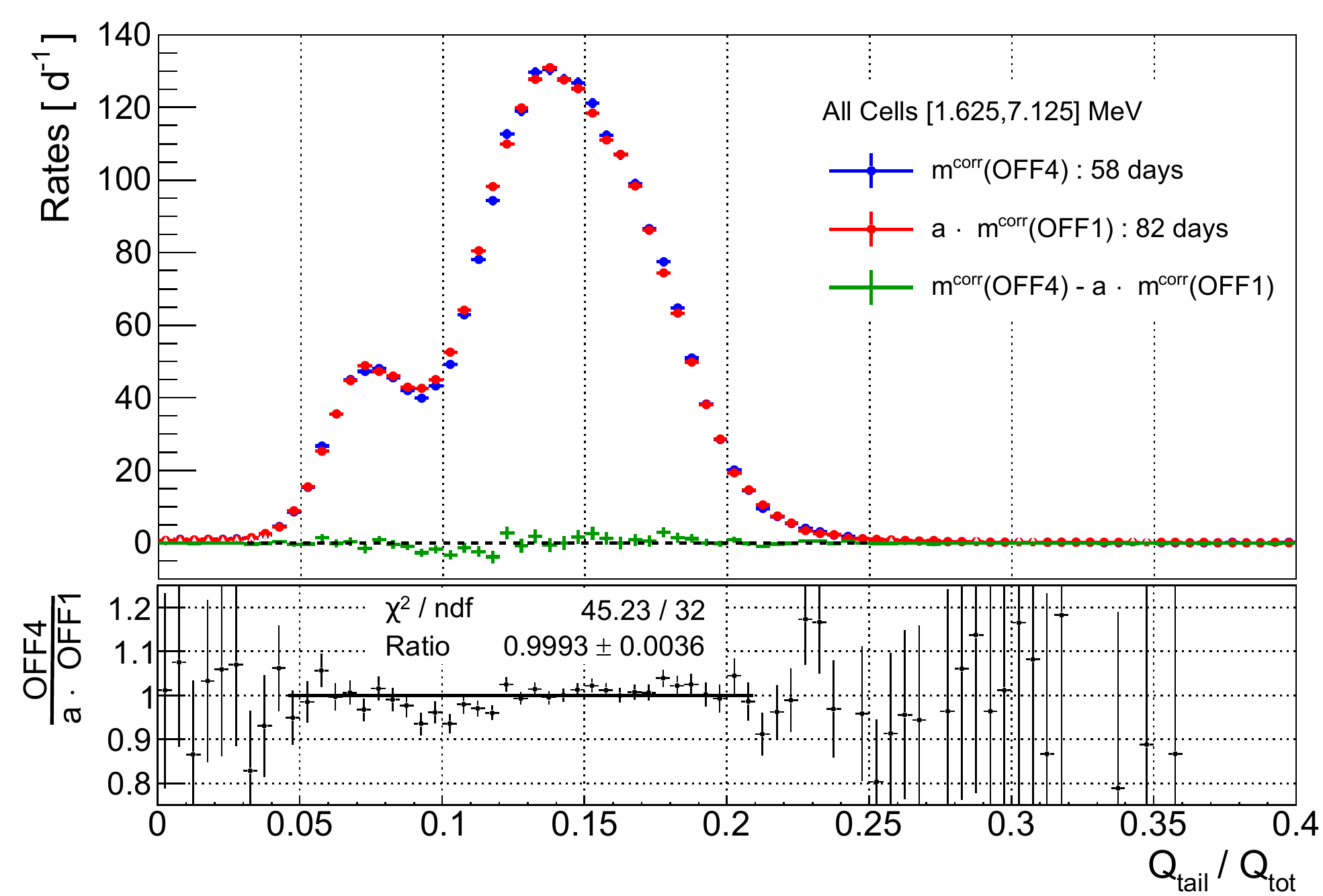}
    \end{indented}
    \caption{OFF1 and OFF4 (early and late phase-II) PSD distributions of correlated events, for all cells and all prompt energy bins in [1.625~MeV, 7.125~MeV]. The scaling parameter $a$ is determined from the PSD fit procedure with the antineutrino component set to zero. A slight shape discrepancy is observed for PSD values around 0.1.}
    \label{fig:OFF1-OFF4}
\end{figure}

\textit{Detector time stability.} The data taking consists of four periods of background measurement when reactor is off (named OFF1 to OFF4) with three reactor-on periods in between (named ON1 to ON3). We found a slight discrepancy in PSD distributions from reactor-off data when comparing extreme time periods: early (OFF1) vs. late (OFF4) phase-II data (figure~\ref{fig:OFF1-OFF4}). First studies indicate that this difference, most visible between proton and electron recoils peaks, might be related to a change in the distance between those peaks or to top/bottom asymmetry in reconstructed energies.

\begin{figure}[tb]
    \begin{indented}
    \item[] \includegraphics[width=0.9\linewidth]{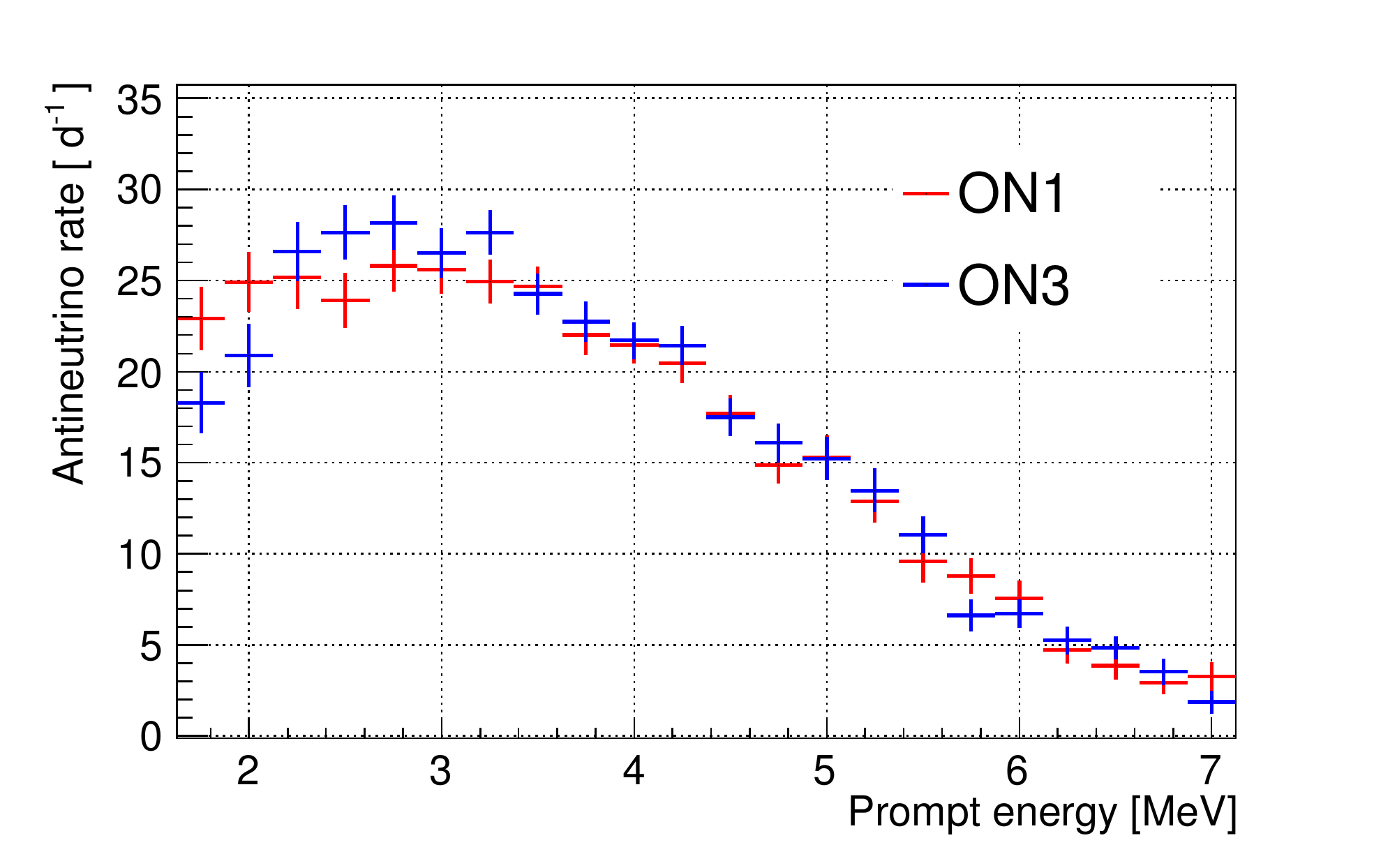}
    \end{indented}
    \caption{Comparison of antineutrino rates extracted from beginning (ON1) and end (ON3) of phase-II data.}
    \label{fig:on1-on3}
\end{figure}

Such reactor-off PSD distributions are used to model correlated background events in reactor-on data. At first order, these drifts are not problematic since linear evolutions are compensated thanks to the alternation of reactor-on and -off periods, but small residual effects may remain. Since the exact mechanism of the drift is not known, we then use the following conservative approach: we compare antineutrino rates from two data sets (beginning and end of phase-II), and treat all statistical tension as potential indication for systematics. Data are split such that the interleaved structure of reactor-on and -off periods is conserved, with reactor-off periods before and after reactor-on: OFF1~- ON1~- OFF2 on one hand, OFF3~- ON3~- OFF4 on the other hand. Resulting antineutrino rates are shown in figure~\ref{fig:on1-on3}.

Although we found a reasonable statistical agreement between early and late phase-II data ($\chi^2/\mathrm{ndf} = 23.2/22$), we use this statistical test to evaluate a potential systematic error related to time evolution of the detector. A systematic uncertainty, proportional to the rate of correlated background events from figure~\ref{fig:SBR}, is added to ON3 data (end of phase-II) to improve the agreement with ON1 data. A $2.0\%$ uncertainty on this background rate is required to retrieve $\chi^2/\mathrm{ndf} = 1$, and taken as systematic uncertainty. In order to allow for distortions in the background spectrum shape, energy bins are assumed to vary independently.\\

As the simulation used to build the response matrix may not reproduce perfectly the detector's response, a second set of systematic uncertainties is related to distortion of the response matrix ($E_\nu \rightarrow E_\mathrm{pr}$ assignment and/or efficiencies may be affected). Below, $R_{ij}^0$ denotes the undistorted response matrix.

\textit{Energy scale.} Uncertainties on the energy scale, namely on the ratio of reconstructed energies $E_\mathrm{rec}^\mathrm{Data} / E_\mathrm{rec}^\mathrm{MC}$, have been described in section~\ref{scn:EScale}. They amount to $1.02/\sqrt{6}\,\%$ (Mn anchoring and energy reconstruction combined) and $0.3\,\%$ (time stability), at the target level. The quadratic sum leads to an overall 0.5\,\% uncertainty. 

To evaluate the impact of this uncertainty on the predicted prompt spectrum, we build another simulation and response matrix $R_{ij}^{+1\sigma}$ with energy scale corresponding to $E^\mathrm{rec,tot}_\mathrm{MC} + 0.5\,\%$ (the effect is assumed to be symetrical). The distorsion of the response matrix, described using a nuisance parameter $\alpha_\mathrm{ES}$ following a standard normal distribution $\mathcal{N}(0,1)$, writes\begin{equation}
    R_{ij} (\alpha_\mathrm{ES}) = R_{ij}^0 + \alpha_\mathrm{ES} \cdot \delta R_{ij}, \qquad \mathrm{with} \quad \delta R_{ij} = \big( R_{ij}^{+1\sigma} - R_{ij}^0 \big).
\end{equation} The corresponding prompt prediction is distorted as $
    N_{j} (\alpha_\mathrm{ES}) = \sum_i R_{ij} (\alpha_\mathrm{ES})\, \Phi_i^{\mathrm{tot},0}$ with full bin-to-bin correlations.

\textit{Selection cuts.} Antineutrino identification versus background rejection is done using the set of cuts described in table \ref{tab:selectionCuts}. Cuts applied on the total reconstructed energy (\#1-2) are taken into account by the systematic uncertainty on the energy scale, but other cuts rely on the reconstructed energy in individual cells, either directly or indirectly. The values of these cuts have been chosen to maximize background rejection, while staying in a region where acceptance is almost constant. Sensitivity of antineutrino rates to cut uncertainties (\textit{i.e.} cut efficiency errors) has been studied taking into account the correlation between cuts. The \STEREO{} simulation was used to generate one pseudo-experiment, then large number of analyses have been done fluctuating the reconstructed energy per cell within their errors. These errors are defined by comparing the reconstructed cell energy to the energy deposit in a cell volume. Resulting antineutrino rates, compared to the rates obtained with the nominal energy reconstruction, allow us to estimate the efficiency cut uncertainties for each cell and each energy bin. For the purpose of the shape analysis, errors have been evaluated summing antineutrino rates from the six cells. This uncertainty $\delta N_j$, correlated between energy bins, is described using a single nuisance parameter $\alpha_\mathrm{Cuts} \sim \mathcal{N}(0,1)$ as $N_j(\alpha_\mathrm{Cuts}) =  N_j^0( 1+\alpha_\mathrm{Cuts} \; \delta N_j)$ with $N_j^0 = \sum_i R_{ij}^0 \Phi_i^{\mathrm{tot},0}$ the nominal prompt spectrum. Consequently, the distortion of the response is written \begin{equation}
   R_{ij} (\alpha_\mathrm{Cuts}) = R_{ij}^0 ( 1+\alpha_\mathrm{Cuts} \; \delta N_j).
\end{equation} 

\textit{Other sources?} We also checked for other potential sources of distortion of the response. First, we studied the impact of having a finite-sized Monte-Carlo sample to generate the response matrix.  To see any change due to slightly different responses, the sample was divided in two equally-sized sets of events from each of which a response matrix $R_{ij}^{(k)}$ was built $(k=1,2)$. Starting from the same toy spectrum $N_j$ and each of these matrices, we extracted antineutrino spectra $\hat{\Phi}^{(k)}$ using fitting frameworks described in section~\ref{scn:Framework}. $\hat{\Phi}^{(1)}$ and $\hat{\Phi}^{(2)}$ were compared using a $\chi^2$ metric: using $10^4$ fluctuating toy spectra, we got on average $\chi^2/\mathrm{ndf} \sim 10^{-3}$. Monte-Carlo samples are thus large enough to not induce any significant change in the fitted antineutrino spectrum.

Second, we checked whether the response matrix actually depends on the nominal IBD yield prediction $\Phi_M^\mathrm{tot}$ used to weight events. To do this, we used two flux models $M \in \{\mathrm{HM}, \mathrm{SM}\}$ to create a response matrix $R_{ij}^{(M)}$ from each. Similarly to the previous study, we extracted antineutrino spectra $\hat{\Phi}^{(M)}$ and compared them with a $\chi^2$ metric: using $10^4$ realizations, we got on average $\chi^2/\mathrm{ndf} \sim 10^{-4}$. The choice of nominal model has then negligible impact as well on the fitted antineutrino spectrum.\\

A final systematic uncertainty comes from the normalization of flux corrections, introduced in section \ref{scn:Prediction}. It is required to extract the \U{}~component of the spectrum. A 5\,\% (resp. 30\,\%) uncertainty is taken for the correction from activated materials (resp. off-equilibrium). The IBD spectrum $\Phi_i^\mathrm{corr}$ induced by flux corrections is thus written as  $\Phi_i^\mathrm{corr} (\alpha_\phi) = \Phi_i^\mathrm{corr} + \alpha_\phi \cdot \delta \Phi_i^\mathrm{corr}$ with a nuisance parameter $\alpha_\phi$ describing the overall normalization uncertainty $\delta \Phi_i^\mathrm{corr}$ of these corrections. \\

\begin{figure}[tb]
    \begin{indented}
	\item[]	\includegraphics[width=0.9\linewidth]{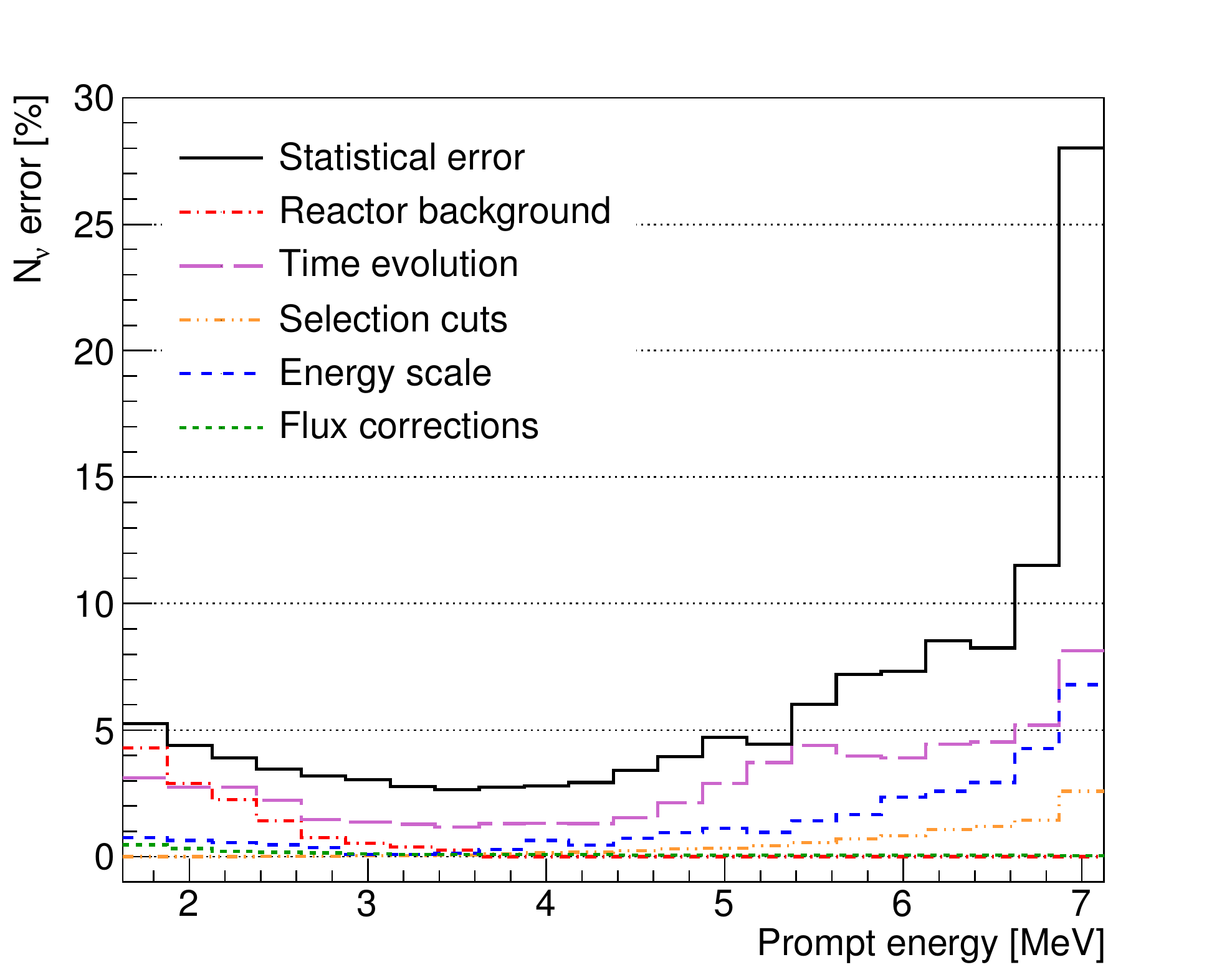}
    \end{indented}
    \caption{Summary of all uncertainties considered for the spectrum shape unfolding.}
    \label{fig:error-summary}
\end{figure}

In summary, the relevant sources of systematic uncertainties are the following: background effects (reactor-induced background and time evolution), impacting the measured antineutrino rate; energy scale and selection cuts, distorting the response matrix; normalization uncertainty of reactor-related flux corrections. The impact of these systematic uncertainties on the prompt spectrum is illustrated in figure~\ref{fig:error-summary}. Normalization uncertainties (cf. \cite[table I]{ratePaper}) are not included since we focus here on spectrum shape analysis. With the \STEREO{} phase-II data presented in this article, statistical uncertainties remain dominant across the energy range.


\section{\label{scn:Framework}Fitting frameworks}

The \U{}~antineutrino spectrum is fitted through the response matrix against \STEREO{} data, using a fitting framework introduced in this section. More specifically, two independent fitting frameworks have been developed in order to compare their outputs and cross-validate this analysis. The main framework is built with nuisance parameters and is used to provide the final result (see section~\ref{scn:results}). The other is built from the experimental covariance matrix and will be briefly introduced as well. 

\subsection{Description}

\textit{First framework: fit with nuisance parameters.} The \U-induced spectrum $\Phi_{M}^\mathrm{U5}$, abbreviated from now on to $\Phi$, is modelled using 20 weights $\lambda_i$ (one per antineutrino energy bin, cf. section~\ref{scn:ResponseMatrix}) describing the deviation to some reference spectrum $\Phi^0$ (also called \textit{prior} in the following): \begin{equation}
    \Phi_i (\vec{\lambda}) = \lambda_i \Phi_i^0.
\end{equation}
Below, unless otherwise specified, we will use the IBD yield prediction from the HM model ($\Phi_{\rm HM}^\mathrm{U5}$) as prior. Taking into account reactor-related flux corrections $\Phi_i^\mathrm{corr}(\vec{\alpha})$, the total IBD spectrum from equation~(\ref{eqn:flux-model}) writes \begin{equation}
    \Phi_i^\mathrm{tot} (\vec{\lambda};\vec{\alpha}) =  \Phi_i (\vec{\lambda}) + \Phi_i^\mathrm{corr}(\vec{\alpha})
\end{equation} and yields the following prompt prediction \begin{equation}
    N_j (\vec{\lambda};\vec{\alpha})  = \sum_i R_{ij} (\vec{\alpha}) \, \Phi_i^\mathrm{tot} (\vec{\lambda};\vec{\alpha}),
\end{equation} where $\vec{\alpha} = (\alpha_\phi, \alpha_\mathrm{ES}, \alpha_\mathrm{Cuts})$ is the set of nuisance parameters described in section~\ref{scn:Systematics}. Each $\alpha_s$ follows a normal law $\mathcal{N}(0,1)$. The best-fit \U{}~spectrum $\hat{\Phi}$, or equivalently the best-fit value $\hat{\lambda}_i$ of $\lambda_i$ parameters, is obtained from the numerical minimization of the following $\chi^2$: \begin{equation}
    \chi^2 (\vec{\lambda};\vec{\alpha}) = \sum_i \left( \frac{N_j(\vec{\lambda};\vec{\alpha}) - D_j}{\sigma_j}   \right)^2 + |\vec{\alpha}|^2 + \mathcal{R}_1(\vec{\lambda})
\end{equation} with $D_j$ the data spectrum in prompt energy space and $\sigma_j$ the associated statistical uncertainty (including background systematics: reactor background, time evolution; see section~\ref{scn:Systematics}). The pull term $|\vec{\alpha}|^2$ controls variations of the nuisance parameters. Finally, $\mathcal{R}_1(\vec{\lambda})$ is a penalty term, or regularization term, that constrains the smoothness of the fitted \U{}~spectrum.

This regularization term is chosen to be the discrete first derivative of $\vec{\lambda}$, with some tunable strength $r>0$: \begin{equation}\label{eqn:reg-1storder}
    \mathcal{R}_1(\vec{\lambda}) = r \sum_{i = 1}^{19} (\lambda_{i+1} - \lambda_i)^2.
\end{equation}
This term constrains the shape of the fitted spectrum around a reasonably smooth shape (encoded in the prior $\Phi^0$). On one hand, it will exclude spectra with large bin-to-bin fluctuations. Such fluctuations in true energy space would be smeared by the resolution of the experiment encoded in the response, so that the shape of $\hat{\Phi}$ may be irregular while the shape of the fitted prompt prediction $\hat{N}_j \sim R\hat{\Phi} $ remains smooth. On the other hand, we have to be careful when choosing the regularization strength $r$. Large values of $r$ will smooth very strongly the fitted spectrum. In the limit $r\rightarrow \infty$, the $\lambda_i$ will become equal for all $i$, \textit{i.e.} the shape of $\hat{\Phi}$ will be the shape of the prior $\Phi^0$.

\begin{figure}[tb]
    \begin{indented}
	\item[]	\includegraphics[width=0.9\linewidth]{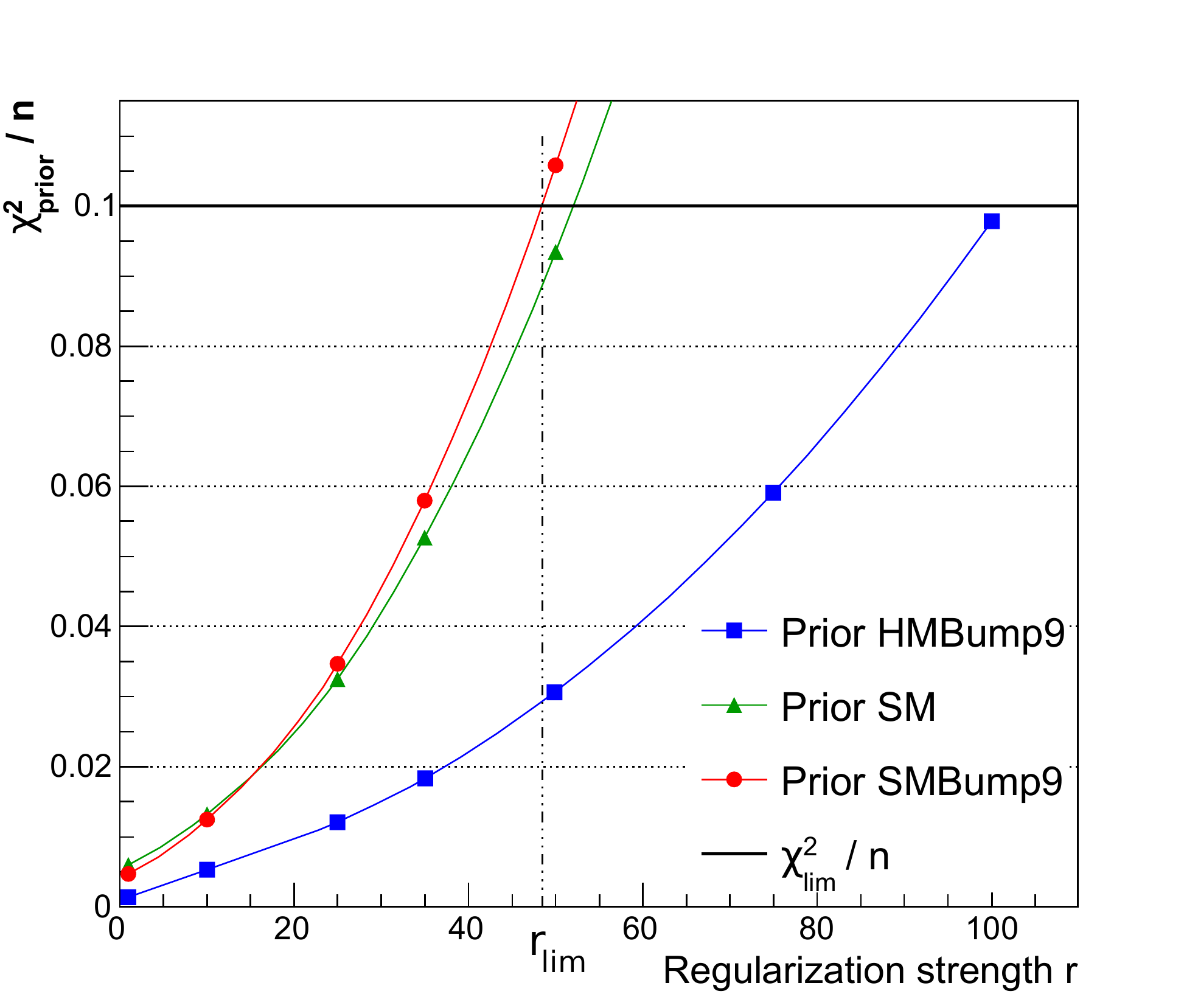}
    \end{indented}
    \caption{Evolution of $\chi^2_\mathrm{prior}/n$ as a function of $r$, from a given toy spectrum drawn around the \texttt{HMBump9} prompt prediction, for several alternate priors. $n=18$ is the number of bins in the analysis range. Largest biases arise when using \texttt{SMBump9} as a prior, which we expect, its shape being the most discrepant w.r.t. nominal \texttt{HM}. In this example, the regularization strength will be taken at the intercept with $\chi^2_\mathrm{lim}/n$, leading to $r_\mathrm{lim} \simeq 48$.}
    \label{fig:bias-prior}
\end{figure}

In order to find a suitable $r$, we study how changing the prior $\Phi^0$ affects the fitted antineutrino spectrum. Indeed, we want to extract a spectrum in antineutrino energy that does not depend on the initial prior's shape. The idea is to compare antineutrino spectra, fitted from the same data set but using the different priors. The nominal prior is built from HM model; alternative priors are built from SM and/or bumped models (see definitions on table~\ref{tab:models}). 
Based on data collected by other experiments, this set of priors should cover the range of plausible antineutrino spectra. As the regularization tends to transfer the prior's shape to the fitted spectrum, the difference between spectra fitted with different priors is expected to grow with $r$. A maximal allowed difference $\chi^2_\mathrm{lim}$ is set, whose value is chosen below. For this study we proceed as follows:\begin{enumerate}
    \item We generate a toy spectrum $D_j$ in prompt space;
    \item From this toy spectrum, we extract the \U{}~spectrum $\hat{\Phi}^{(1)}$ using the nominal prior, as well as $\hat{\Phi}^{(2)}$ using any alternative prior;
    \item The difference is evaluated as \begin{equation}
        \chi^2_\mathrm{prior} = \sum_{ii'} \left( \hat{\Phi}_i^{(1)} - \hat{\Phi}_i^{(2)} \right) V_{\hat{\Phi},ii'}^{-1} \left( \hat{\Phi}_{i'}^{(1)} - \hat{\Phi}_{i'}^{(2)} \right)
        \end{equation} 
    with $V_{\hat{\Phi}}$ the covariance matrix of the nominal $\hat{\Phi}^{(1)}$ antineutrino spectrum, restricted to the analysis range. $V_{\hat{\Phi}}$ is built, for any value of $r$, by sampling: $10^4$ toy spectra $D_j$ are drawn in prompt space and the corresponding best-fit antineutrino spectra $\hat{\Phi}$ are extracted; $V_{\hat{\Phi}}$ results from their distribution. An example of the evolution of $\chi^2_\mathrm{prior}$ as a function of $r$ is displayed in figure \ref{fig:bias-prior};
    \item We find $r_\mathrm{lim}$, the largest value of $r$ such that, for \textit{all} alternative priors, we have $\chi^2_\mathrm{prior} \leqslant \chi^2_\mathrm{lim}$.
\end{enumerate}
This procedure is repeated for thousands of toy spectra distributed around several prompt predictions (based on \texttt{HM}, \texttt{SM}, and bumped models \texttt{HMBump9}, \texttt{SMBump9}). 

\begin{figure}[tb]
    \begin{indented}
		\item[]\includegraphics[width=0.9\linewidth]{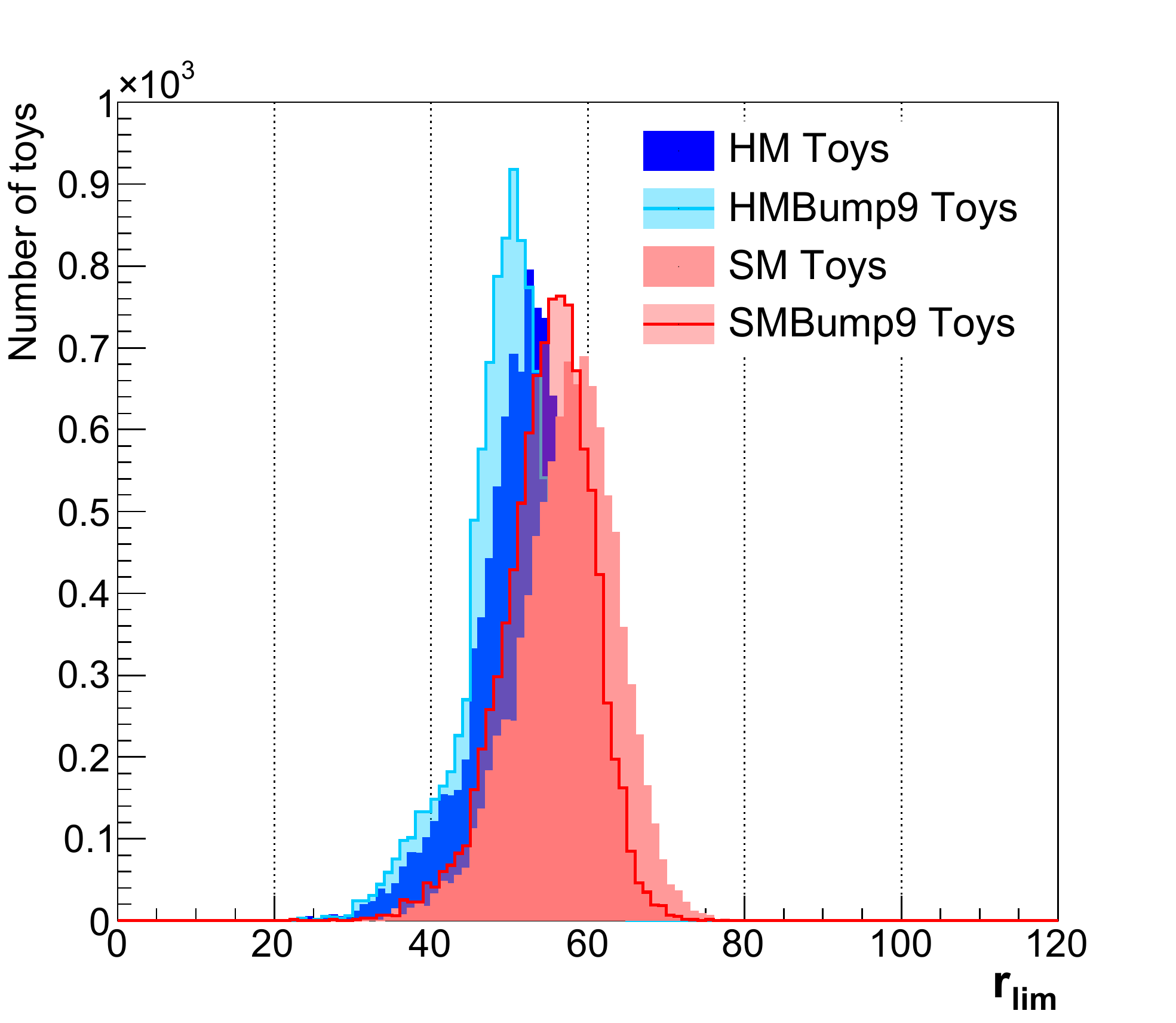}
    \end{indented}
    \caption{Distribution of $r_\mathrm{lim}$ for toy spectra drawn around several prompt predictions (based on \texttt{HM}, \texttt{SM}, \texttt{HMBump9} or \texttt{SMBump9} models). For all toys, the reference spectrum is \texttt{HM}. The average value over all toy spectra, $\langle r_\mathrm{lim} \rangle =53$, will be used when fitting actual data.}
    \label{fig:rlim}
\end{figure}

In order to define an appropriate value for $\chi^2_\mathrm{lim}$, we consider two aspects. First, $\Phi^{(1)}$ and $\Phi^{(2)}$ do not differ by a statistical fluctuation but only by a different regularization. Denoting $n$ the number of bins in the analysis range, we then expect $\chi^2_\mathrm{prior}/n$ to be (much) smaller than 1, which is guaranteed if $\chi^2_\mathrm{lim}/n$ is also (much) smaller than 1. Second, $\chi^2_\mathrm{prior}/n$ is a measure of the bias introduced by a change of prior -- specifically (bias/error)$^2$ -- that we may call \textit{regularization error}. In case this error is negligible with respect to the covariance $V_{\hat{\Phi}}$ of the unfolded spectrum, it needs not to be taken into account any further. When the regularization error is added in quadrature to the original error, we get \begin{equation} 
    \mathrm{error}'   = \sqrt{ \mathrm{error}^2 + \mathrm{bias}^2 } \simeq \mathrm{error} \times \big[ 1+ (\mathrm{bias/error})^2/2 \big].
  \end{equation}
We then need $\chi^2_\mathrm{prior}/2n \ll 1$. Consequently we chose in this analysis to take $\chi^2_\mathrm{lim} / n = 0.1$.

The distribution of $r_\mathrm{lim}$ obtained with this value of $\chi^2_\mathrm{lim}$ is shown in figure~\ref{fig:rlim}. The average value, $\langle r_\mathrm{lim} \rangle =53$, will be used to fit actual \STEREO{} data in section~\ref{scn:results}.\\

\noindent\textit{Second framework: fit with covariance matrix.} The second framework allows to extract the antineutrino spectrum by minimizing a $\chi^2$ written in a covariance matrix approach. In this regard, we denote $V_\mathrm{pr}$ the $22 \times 22$ experimental covariance matrix of the measured spectrum in prompt space. Using the same notations introduced before ($D$ is the data vector, $\Phi$ and $\Phi^\mathrm{corr}$ the \U{}~spectrum and reactor-related corrections, $R$ the response matrix), the $\chi^2$ to be minimized reads:
\begin{equation}
\chi^2 (\Phi)  = \Big(D - R (\Phi+\Phi^\mathrm{corr}) \Big)^{T}{V_\mathrm{pr}^{-1}}\Big(D - R (\Phi+\Phi^\mathrm{corr}) \Big) \; + \mathcal{R}_1 (\vec{\lambda}). \label{eq:chi2}
\end{equation}
The second term is the same first-order regularization term as for the first framework (cf. equation~(\ref{eqn:reg-1storder})). In this framework, $r = 30$ is set based on the Generalized Cross-Validation (GCV) prescription \cite{GCV}. 

The minimization of the $\chi^2$ defined in equation~(\ref{eq:chi2}) is done analytically. Let $M_1$ be the matrix so that $ \lambda_{i+1} - \lambda_{i} = [M_1 \cdot\Phi]_{i}$. The unfolded spectrum and its covariance matrix then read:
 \begin{equation}
\hat{\Phi} = H \cdot  D \; , \qquad V_{\hat{\Phi}} =H \, V_\mathrm{pr} H^{T}
\end{equation}
where $H =\left(R^T V_\mathrm{pr}^{-1} R + r \, M_1^T M_1 \right)^{-1} R^T V_\mathrm{pr}^{-1}$. \\

\subsection{Validations}
\noindent\textit{Bias studies.}
To assess performances of the frameworks described above, bias studies are performed. For a given \U{}~prediction $\overline{\Phi}$: \begin{enumerate}
\item The model is folded to prompt space using the response matrix;
    \item We generate $10^4$ fluctuated toy spectra $D_j$ in prompt space;
    \item From each toy spectrum, we extract the best-fit spectrum $\hat{\Phi}$;
    \item The average spectrum $\langle \hat{\Phi} \rangle$ is compared to the initial value of the prediction $\overline{\Phi}$. 
\end{enumerate}
Results are given below for the first framework.

When $\overline{\Phi} $ is the \texttt{HM} model ($\Phi_\mathrm{HM}^\mathrm{U5}$), it has the same shape than the prior $\Phi^0$. Hence the regularization term imposes the correct shape to the fitted spectra, and the relative bias $(\langle \hat{\Phi}_i \rangle - \overline{\Phi}_i)/\overline{\Phi}_i$ is not larger than $0.1\,\%$.

\begin{figure}[tb]
    \begin{indented}
	\item[]	\includegraphics[width=0.9\linewidth]{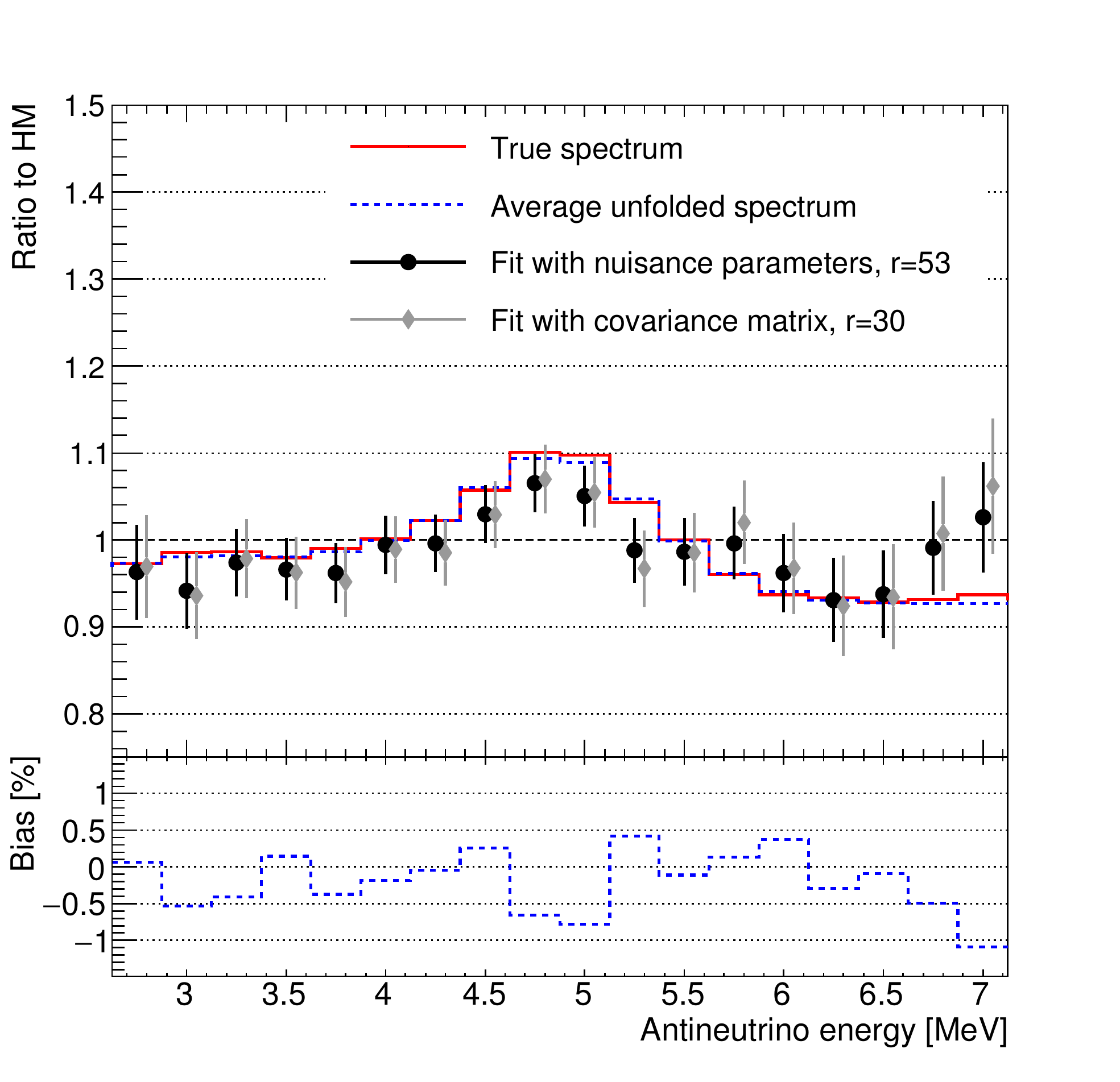}
    \end{indented}
    \caption{(Top) One example of unfolded best-fit spectra $\hat{\Phi}$, extracted using the fit with nuisance parameters or the fit with covariance matrix, from the same fluctuated toy. The true model is built by adding a 12\,\% bump on \texttt{SM} prediction, with $(\mu,\sigma) = (4.8~\mathrm{MeV},0.35~\mathrm{MeV})$. The average unfolded spectrum is computed using $10^4$ fluctuated toys, individually unfolded. (Bottom) Average bias for this specific toy model.}
    \label{fig:fakedata}
\end{figure}

When $\overline{\Phi}$ is a bumped model (see an example on figure~\ref{fig:fakedata}), we retrieve very well the bump on the mean spectrum $\langle \hat{\Phi} \rangle$, with sub-percent biases. One can notice that the unfolded bump is however slighlty flattened by the regularization, which constrains the spectrum shape to be similar to the un-bumped nominal prior (\texttt{HM}). This effect is small for bumps with similar amplitude and width as what we observe in \STEREO{} data (cf. section~\ref{scn:results}), but biases do become significant for sharper bumps.

Finally, when we use the Summation Model ($\Phi_\mathrm{SM}^\mathrm{U5}$) for $\overline{\Phi}$, the relative bias remains smaller than $1\,\%$ in the analysis range. Though \texttt{SM} and the prior \texttt{HM} have different shapes, the regularization strength is flexible enough to correctly recover the initial model.

We conclude that our binning and regularization procedure are suitable to study local distortions to the HM model similar to what was observed by other experiments. \\

\noindent\textit{Comparison of the two frameworks.} Having two independent frameworks allows to compare their output and validate the fitting procedure. Let us emphasize their complementarity here again: nuisance parameters vs. experimental covariance matrix; numerical vs. analytical minimization; choice of $r$ using physical interpretation (in terms of bias) vs. using an established statistical criterion (GCV). Finding similar best-fit spectra will then confirm the robustness of this analysis.

The comparison is done using one of the fluctuated toy spectrum associated with the bumped model on figure~\ref{fig:fakedata}. This fluctuated toy spectrum is fitted by both frameworks and best-fit \U{}~spectra are compared: they are in a good agreement with the true one ($\chi^2/\mathrm{nbins} = 13.6/18$ and $\chi^2/\mathrm{nbins} = 19.8/18$). Deviations observed in some bins are related to statistical fluctuations; on average, biases remain small (<1\%, see bottom panel of figure~\ref{fig:fakedata}). The GCV prescription leads to a slightly weaker regularization with $r=30$; as a result, the spectrum from the second framework has slightly larger variance. The $\chi^2$ agreement between spectra fitted with the first and second framework is $\chi^2/\mathrm{nbins} = 0.4/18$. Since spectra originate from the same statistical fluctuation, it is expected to find $\chi^2/\mathrm{nbins} \ll 1$. We also tested a second order regularization term \begin{equation}    \mathcal{R}_2 (\vec{\lambda})  = r \sum_i \left( \lambda_{i+1} - 2 \lambda_i + \lambda_{i-1} \right)^2\end{equation} in the covariance matrix framework for this comparison, and obtained comparable results.

Having fully validated the fit frameworks, we will use the first one, built with nuisance parameters and a carefully tuned regularization strength, to provide \STEREO{} unfolded spectrum and covariance matrix in the next section.


\section{\label{scn:results}Results and discussion}

\begin{figure}[tb]
    \begin{indented}
	\item[]	\includegraphics[width=\linewidth]{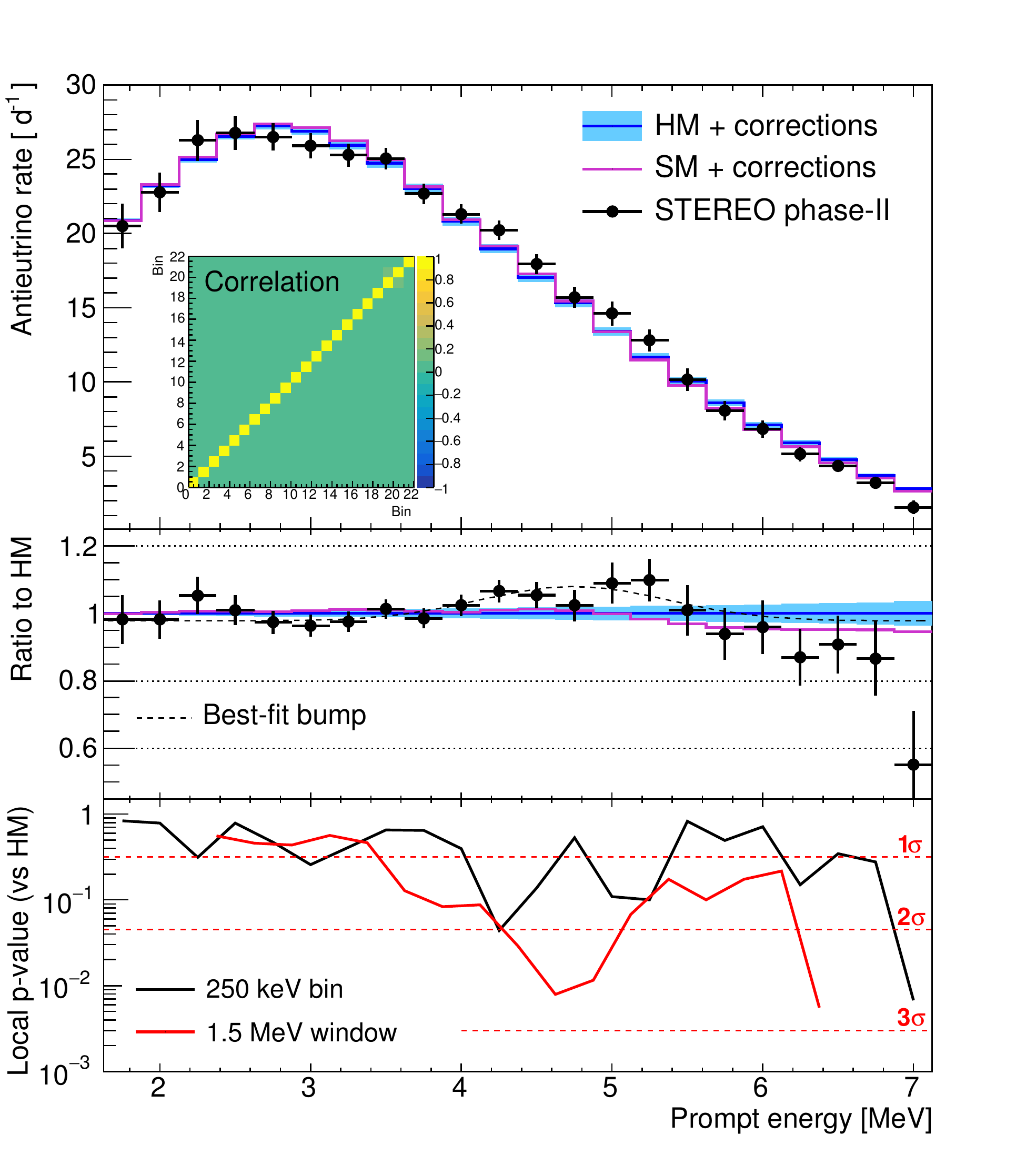}
    \end{indented}
    \caption{(Top) Measured IBD yield spectrum along with area-normalized Huber-Mueller (HM) and Summation model (SM) predictions, including reactor-related corrections. Data error bars include statistical and systematic uncertainties. The almost diagonal correlation matrix is displayed. The blue error band on the HM prediction include theoretical uncertainties from \cite{Huber:2011wv} without the normalization component. (Middle) Ratios to HM prediction. (Bottom) Local $p$-value quantifying the significance of deviations from HM for each individual 250 keV bin and for a 1.5~MeV sliding window (6 consecutive bins).}
    \label{fig:resultprompt}
\end{figure}

\begin{figure}[tb]
    \begin{indented}
	\item[]	\includegraphics[width=\linewidth]{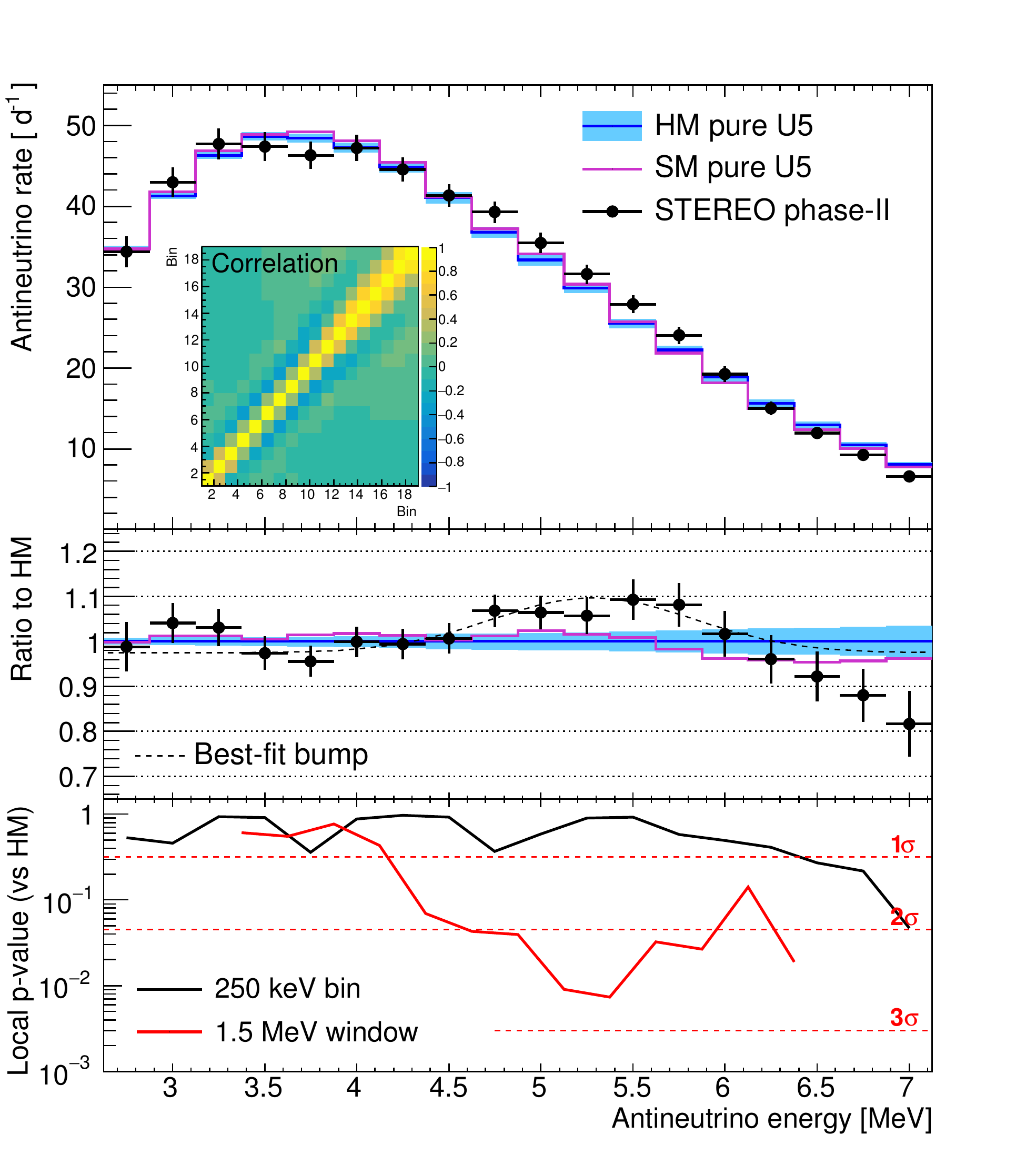}
    \end{indented}
    \caption{(Top) Unfolded \U{}~IBD spectrum along with area-normalized Huber-Mueller (HM) and Summation Model (SM) predictions. Data error bars are taken from diagonal coefficients of the total covariance matrix and include statistical and systematic uncertainties. The non-trivial correlation matrix is displayed. Model errors from \cite{Huber:2011wv}, without the normalization component, are shown as the blue error band for reference only. (Middle) Ratios to HM prediction. (Bottom) Local $p$-value quantifying the significance of deviations from HM for each individual 250 keV bin and for a 1.5~MeV sliding window (6 consecutive bins).}
    \label{fig:resulttrue}
\end{figure}

The measured IBD spectrum from \STEREO{} phase-II is shown in figure \ref{fig:resultprompt} (top panel). A $\chi^2$ test is performed as \begin{equation}\label{eqn:chi2test-prompt}
    \chi^2(\mathrm{data~vs.~model}\,M) = \sum_{jj'}( D_j -  M_j) \left[V_\mathrm{pr}^{-1} \right]_{jj'} ( D_{j'} -  M_{j'}) 
\end{equation} with $V_\mathrm{pr}$ the experimental covariance matrix in prompt energy and $M_j = \sum_i R_{ij}^0 \, \Phi_{M,i}^\mathrm{tot}$ the prompt prediction resulting from model $M$. It gives an agreement of $\chi^2/\mathrm{ndf} = 28.4/22$ against the HM \cite{Huber:2011wv} prediction and $\chi^2/\mathrm{ndf} = 23.8/22$ against the SM \cite{Estienne:2019ujo} prediction. For this test, both predictions are area-normalized to the measured spectrum and include reactor-related corrections (activation and off-equilibrium). Model uncertainties are not taken into account for the results presented in the following. Uncertainties of the Huber-Mueller prediction \cite{Huber:2011wv} are dominated by a normalization effect and not required for a shape-only analysis. For reference, the shape uncertainties are shown as a blue band in figures~\ref{fig:resultprompt} and \ref{fig:resulttrue}. As for the Summation Model, uncertainties are not provided in the publication \cite{Estienne:2019ujo}.
    
Local significance of deviations between data and the HM prediction are computed using the same method as in \cite{Ashenfelter:2018jrx,An:2016srz} for a 250 keV or 1.5 MeV window~$W$ (1 or 6 consecutive bins, respectively). The 1.5 MeV window is relevant to look at structures such as the bump observed by other experiments. Using the same method also allows us to compare the measured significance of distortions with the above-mentioned publications. The method consists in adding a free parameter in the $\chi^2$ for each bin of interest (1 or 6) to allow the data points to float towards the model: $\forall j \in W,\; D_j \rightarrow \alpha_j \,D_j$. The $\chi^2$ test is performed as in eq.~(\ref{eqn:chi2test-prompt}), treating all $\alpha$ parameters as nuisance, and gives a new best fit $\chi^2_{\mathrm{new},W}$ (obviously smaller than the initial one). The $\Delta \chi^2_W = \chi^2 - \chi^2_{\mathrm{new},W}$ quantity states how much worse is the agreement to the model when data points of the window $W$ are at their measured value ($\chi^2$) instead of free-floating ($\chi^2_{\mathrm{new},W}$): this estimates the impact of the window of interest on the overall agreement. A local $p$-value is computed from $\Delta \chi^2_W$ assuming 1 or 6 degrees of freedom (bottom panel of figure \ref{fig:resultprompt}). Two regions reach the $2\sigma$ threshold of local $p$-value: an excess of events around 4.5-5 MeV and a deficit at the high-energy end of the spectrum. The high-energy deficit is particularly significant for the last energy bin (centered at 7 MeV) with more than 2.5 standard deviations with respect to the HM prediction. As discussed in section~\ref{scn:Data}, specific investigations indicate that a large part of this deficit might be due to a downward statistical fluctuation enhanced by the PSD fit procedure.

The unfolded pure-\U{}~IBD yield spectrum $\hat{\Phi}$ is shown in figure \ref{fig:resulttrue}. The same $\chi^2$ test as equation (\ref{eqn:chi2test-prompt}) is performed, restricted to the analysis range [2.625 MeV, 7.125 MeV], using the prediction $M_j = \Phi_{M,j}^\mathrm{U5}$ and the covariance $V_{\hat{\Phi}}$ in antineutrino energy. It gives an agreement of $\chi^2/\mathrm{ndf} = 26.7/18$ against the HM prediction and $\chi^2/\mathrm{ndf} = 20.6/18$ against the SM prediction. Here reactor-related corrections have been applied and model predictions are area-normalized to the unfolded data spectrum. Again, theoretical uncertainties are not included. Local significances are computed as in prompt energy, for a single bin and a 1.5~MeV sliding window. Due to non-negligible bin-to-bin correlations from the fitting framework the significance of a single bin's deviation rarely crosses the 1$\sigma$ level. As for prompt energy, local $p$-value of the 1.5~MeV sliding window however shows two regions of more significant ($>2\sigma$) deviations to the HM model: around 5.5 MeV and centered at 7 MeV. The high energy deficit, driven mostly by the last prompt energy bin on figure~\ref{fig:resultprompt}, spreads on several consecutive $E_\nu$ bins. This is related to correlations introduced in the spectrum by the response matrix (and the regularization term). Using a 1.5 MeV sliding window, the local significance in prompt and antineutrino energy are, as expected, quantitatively similar at $\sim 2.5\sigma$. 

In order to further quantify the excess in the 5~MeV region, and to allow comparisons with other published results, a gaussian model is introduced. Equation (\ref{eqn:chi2test-prompt}) is modified to
\begin{equation}\label{eqn:gauss-fit-prompt}
    \chi^2(\boldsymbol{\theta}) = \sum_{jj'}\big( D_j - M_j (\boldsymbol{\theta})\big) \left[V_\mathrm{pr}^{-1} \right]_{jj'} \big( D_{j'} - M_{j'} (\boldsymbol{\theta})\big) 
\end{equation} where the baseline model (HM) is extended with a free normalization and a gaussian bump: $M_j(\boldsymbol{\theta}) = M_j\, \big( \varphi + \mathcal{G}(A, \mu, \sigma) \big)$. Best-fit parameters give an amplitude of $A = 10.1 \pm 2.9~\%$ and a mean prompt energy of $\mu = 4.75 \pm 0.21~\mathrm{MeV}$, which confirms the existence of such a bump with about $3.5\sigma$ significance. The corresponding minimal $\chi^2/\mathrm{nbins} = 15.6/22$ provides, as expected, a much better agreement with data than HM (28.4/22) or SM (23.8/22) predictions. We also investigate how the best-fit bump varies when the baseline model $M_j$ is changed to Summation Model prediction. We obtain similar amplitude and significance: $A = 10.1 \pm 3.1~\%$ ($3.3\sigma$). The central energy $\mu = 4.94 \pm 0.25~\mathrm{MeV}$ is larger but consistent within error bars. Best-fit parameters and goodness of fit are summarized in table~\ref{tab:best-fit-bump}.

Similar gaussian fits in antineutrino energy lead to the same conclusions. With a baseline model following the HM prediction, one obtains an amplitude of $A = 12.1 \pm 3.4~\%$ ($3.5\sigma$) and a mean energy of $\mu = 5.29 \pm 0.18~\mathrm{MeV}$ ($\chi^2/\mathrm{nbins} = 12.3/18$). Comparable amplitude and central value are obtained when starting from the SM prediction. Bump characteristics in prompt and antineutrino energy are consistent with what we expect from the response matrix: higher mean energy on the unfolded spectrum, related to the energy shift from equation (\ref{eqn:Epr_vs_Enu}); lower amplitude and larger width in prompt energy, due to detector smearing. Moreover, the excess observed by \STEREO{} is compatible, though at slighlty lower energy, with the bump observed by the Daya Bay experiment \cite{An:2016srz}, whose central energy was estimated in \cite{andriamirado2020improved} to be at $E_\nu \simeq 5.68~\mathrm{MeV}$.

\begin{table}[tb]
    \caption{Best-fit parameters for the gaussian fit $\mathcal{G}(A,\mu,\sigma)$ of the event excess in \STEREO{} data.}
    \label{tab:best-fit-bump}
    \begin{indented}
    \item[] \begin{tabular}{c||c|c||c|c}
    \hline \hline
    \multirow{2}{*}{Parameter}        & \multicolumn{2}{c||}{Antineutrino energy} & \multicolumn{2}{c}{Prompt energy} \\
            & \texttt{HM} + bump & \texttt{SM} + bump & \texttt{HM} + bump & \texttt{SM} + bump \\
    \hline
    Amplitude $A$ (\%) & $ 12.1 \pm 3.4$&$11.8 \pm 3.6$ & $10.1\pm 2.9$ & $10.1 \pm 3.1$\\
   Mean energy $\mu$ (MeV) & $5.29 \pm 0.18$& $5.47 \pm 0.21$ & $ 4.75\pm 0.21$ & $4.94 \pm 0.25$ \\
   Width $\sigma$ (MeV) & $ 0.55\pm0.17$& $0.60 \pm 0.20$ & $0.63\pm 0.17$& $0.69\pm 0.19$ \\ \hline
   Goodness of fit $\chi^2/\mathrm{nbins}$& 12.3/18 & 9.1/18 & 15.6/22 & 12.5/22\\
    \hline \hline
    \end{tabular}
    \end{indented}
\end{table}

With current phase-II data, \STEREO{} shows the presence of a local excess with respect to HM prediction. Concerning the fuel origin of the 5~MeV excess, the fitted amplitude ($12.1 \pm 3.4~\%$) does not discriminate yet between a pure \U{}~bump ($\sim 15~\%$) and a bump equally shared by all isotopes ($\sim 9~\%$).


\section{\label{scn:conclusion}Conclusion}

In addition to an accurate measurement of the total antineutrino rate from \U{}~fissions \cite{ratePaper}, the \STEREO{} experiment reports its first measurement of the spectral shape using 118 days of reactor-on (phase-II). Special care was given to the control of energy reconstruction and quantification of energy scale distortions compared to the MC detector model, which where found to be very small (<1\,\%). Small time evolution has been observed and a conservative systematic uncertainty has been added.

The spectrum is reported in reconstructed prompt energy as well as in antineutrino energy. The unfolding procedure has been carefully studied so as to not introduce any significant bias in the spectrum determination. Two independent fitting frameworks have been developed and their results, consistent with each other, validate the unfolding process. With phase-II data, \STEREO{} demonstrates its ability to observe and quantify local distortions in the \U{}~spectrum. An excess of events with respect to the Huber-Mueller prediction \cite{Huber:2011wv} is found in the $E_\nu = 5$~-~5.5~MeV range. A gaussian fit of this excess gives an amplitude of $A = 12.1 \pm 3.4~\%$, ruling out the no-bump hypothesis at the $3.5\sigma$ level. A high energy deficit around $E_\nu =7$~MeV is also observed. These local distortions are present both in the prompt energy and the antineutrino energy spectra and do not arise from the unfolding method. This article thus provides a reliable unfolded antineutrino spectrum, the first time for an experiment working with HEU fuel. The latest Summation Model \cite{Estienne:2019ujo} appears to provide better agreement with the observed data though the lack of published uncertainties prevents from any strong statement.

Inclusion of forthcoming \STEREO{} phase-III data will provide a twice larger data set and increase sensitivity to local distortions. A joint unfolding, including data from other experiments \cite{andriamirado2020improved} using highly enriched uranium (HEU) fuel is also on-going, and will further increase the accuracy of such studies.

To allow reproducibility of results, inputs and results of this analysis are provided to the community under the reference \cite{datashare}.


\section*{\label{sec:acknowledgments}Acknowledgements}
We thank G. Mention  for valuable inputs on the unfolding method based on the covariance matrix. This work is funded by the French National Research Agency (ANR) within the project ANR-13-BS05-0007 and the ``Investments for the future'' programme ENIGMASS LabEx (ANR-11-LABX-0012). Authors are grateful for the technical and administrative support of the ILL for the installation and operation of the \STEREO{} detector. We further acknowledge the support of the CEA, the CNRS/IN2P3 and the Max Planck Society.

\nolinenumbers
\section*{References}
\bibliographystyle{unsrt}
\bibliography{references}

\end{document}